\documentclass[times,twocolumn,review]{elsarticle}

\usepackage{medima}
\usepackage{framed,multirow}
\usepackage{amsmath}
\usepackage{amssymb}
\usepackage{tabularx}
\usepackage{multirow}        
\usepackage{booktabs}
\usepackage{latexsym}
\usepackage{float}
\usepackage{appendix}

\usepackage{url}
\usepackage{xcolor}

\usepackage{hyperref}

\definecolor{newcolor}{rgb}{.8,.349,.1}

\journal{Medical Image Analysis}

\begin{document}

\verso{Tianshu Zheng \textit{et~al.}}

\begin{frontmatter}

\title{A microstructure estimation Transformer inspired by sparse representation for diffusion MRI }%

\author[1]{Tianshu \snm{Zheng}}
\author[2]{Cong \snm{Sun}}
\author[3]{Weihao \snm{Zheng}}
\author[1]{Wen \snm{Shi}}
\author[1]{Haotian \snm{Li}}
\author[4]{Yi \snm{Sun}}
\author[1]{Yi \snm{Zhang}}
\author[2]{Guangbin \snm{Wang}}
\author[5]{Chuyang \snm{Ye}}
\author[1]{Dan \snm{Wu} \corref{cor1}}

\cortext[cor1]{Corresponding author: 
  email: danwu.bme@zju.edu.cn}

\address[1]{Department of Biomedical Engineering, College of Biomedical Engineering $\&$ Instrument Science, Zhejiang University, Hangzhou, Zhejiang, China}
\address[2]{Department of Radiology, Shandong Medical Imaging Research Institute, Cheeloo College of Medicine, Shandong University, Jinan, China}
\address[3]{School of Information Science and Engineering, Lanzhou University, Lanzhou, China}
\address[4]{MR Collaboration, Siemens Healthineers Ltd., Shanghai, China}
\address[5]{School of Integrated Circuits and Electronics, Beijing Institute of Technology, Beijing, China}

\received{1 May 2013}
\finalform{10 May 2013}
\accepted{13 May 2013}
\availableonline{15 May 2013}
\communicated{S. Sarkar}

\begin{abstract}
Diffusion magnetic resonance imaging (dMRI) is an important tool in characterizing tissue microstructure based on biophysical models, which are typically multi-compartmental models with mathematically complex and highly non-linear forms. Resolving microstructures from these models with conventional optimization techniques is prone to estimation errors and requires dense sampling in the q-space with a long scan time. Deep learning based approaches have been proposed to overcome these limitations in dMRI-based microstructure estimation. Motivated by the superior performance of the Transformer in feature extraction than the convolutional structure, in this work, we present a learning-based framework based on Transformer, namely, a \emph{Microstructure Estimation Transformer with Sparse Coding} (METSC) for dMRI-based microstructural parameter estimation with downsampled q-space data. To take advantage of the Transformer while addressing its limitation in large training data requirement, we explicitly introduce an inductive bias—model bias into the Transformer using a sparse coding technique to facilitate the training process. Thus, the METSC is composed with three stages, an embedding stage, a sparse representation stage, and a mapping stage. The embedding stage is a Transformer-based structure that encodes the signal in a high-level space to ensure the core voxel of a patch is represented effectively. In the sparse representation stage, a dictionary is constructed by solving a sparse reconstruction problem that unfolds the \emph{Iterative Hard Thresholding} (IHT) process. The mapping stage is essentially a decoder that computes the microstructural parameters from the output of the second stage, based on the weighted sum of normalized dictionary coefficients where the weights are also learned. We tested our framework on two dMRI models with downsampled q-space data, including the \emph{intravoxel incoherent motion} (IVIM) model and the \emph{neurite orientation dispersion and density imaging} (NODDI) model. The proposed method achieved up to 11.25 folds of acceleration in scan time while retaining high fitting accuracy, reducing the \emph{mean square error} (MSE) by up to 70\% compared with q-space learning. METSC outperformed the other state-of-the-art learning-based methods, including the model-free and model-based methods, and reduced the MSE by most 81\%. The network also showed robustness against the noise. The superior performance of METSC indicates its potential to improve dMRI acquisition and model fitting in clinical applications.
\end{abstract}

\begin{keyword}

\KWD Diffusion MRI\sep Microstructural model\sep Sparse coding \sep Transformer
\end{keyword}

\end{frontmatter}


\section{Introduction}

Diffusion MRI (dMRI) is one of the most important medical imaging tools and the only noninvasively approach that can probe tissue microstructures based on the restricted diffusion of water molecules in biological tissues ~\citep{mori2006principles}. The commonly used diffusion tensor model has shown to be sensitive to pathological changes such as stroke and tumor ~\citep{le2001diffusion}, but it is not specific to microstructural properties, such as cell size, axonal diameter, fiber density and orientational dispersion. Advanced dMRI models are developed to characterize specific microstructural features ~\citep{novikov2019quantifying}, such as \emph{intravoxel incoherent motion} (IVIM) ~\citep{le1988separation}, AxCaliber ~\citep{assaf2008axcaliber}, \emph{diffusion basis spectrum imaging} (DBSI) ~\citep{wang2011quantification}, \emph{neurite orientation dispersion and density imaging} (NODDI) ~\citep{zhang2012noddi}, \emph{soma and neurite density imaging} (SANDI) ~\citep{palombo2020sandi}, and \emph{imaging microstructural parameters using limited spectrally edited diffusion} (IMPULSED) ~\citep{jiang2016quantification}, just to name a few. The majority of the advanced dMRI models consist of multiple compartments with mathematically complex and highly non-linear signal representations. Fitting of these models with conventional nonlinear optimization techniques, such as nonlinear least square fitting~\citep{arun1987least}, is prone to estimation errors. Moreover, from the data acquisition perspective, advanced dMRI models require the acquisition of multiple b-values and diffusion directions in the q-space, which is time-consuming and vulnerable to motion artifacts. This is particularly a problem for moving subjects, such as abdominal organs, fetuses, and placentas. 

To reduce the estimation error and accelerate the acquisition for advanced dMRI models, many methods have been proposed. Nedjati-Gilani et al. ~\citep{nedjati2014machine} and Alexander et al. ~\citep{alexander2014image,alexander2017image} proposed a random forest method to estimate the microstructural parameters in the Kärger model ~\citep{karger1988principles}, NODDI model, and \emph{spherical mean technique} (SMT) model ~\citep{kaden2016quantitative}, respectively. The development of deep learning techniques opens a new avenue for dMRI model fitting. The concept of q-space deep learning ~\citep{golkov2016q} is first proposed to directly map the dMRI signals to the DKI parameters using a subset of the q-space data (reduced number of b-value and diffusion directions). The original q-space deep learning (abbreviated as q-DL) only used the three-layer multilayer perceptron (MLP)  ~\citep{golkov2016q}. Gibbons et al. used a 2D convolution neural network to estimate the NODDI and generalized fractional anisotropy maps simultaneously ~\citep{gibbons2019simultaneous}. Koppers et al. used a residual network to increase the comparability of dMRI signals measured on two different scanners ~\citep{koppers2019spherical}. Chen et al. used a subset q-space to estimate the NODDI parameters via graph convolutional neural network ~\citep{chen2020estimating}. Barbieri et al. used three-layer MLP to estimate the IVIM model parameters ~\citep{barbieri2020deep}. Beyond the end-to-end mapping approaches, the model-driven neural networks that introduce domain knowledge into a deep neural network as the prior information have also been proposed to improve network performance and interpretability ~\citep{gregor2010learning,yang2018admm,xu2018model,wang2020model,liang2019deep}. Specifically, the model-driven neural network is designed to unfold the optimization process of a mathematical model through a network ~\citep{liang2019deep}. In contrast to the conventional networks, the model-driven network is not only data-driven, but also incorporates a model prior that makes the network easy to be interpreted ~\citep{wang2020model}, and therefore, gained increasing popularity in the medical image area. Gregor et al. first proposed a sparse coding neural network based on the optimization procedure ~\citep{gregor2010learning}. ADMM-Net is one of the commonly used model-driven deep neural networks and was first used in MRI for solving the compressed sensing problem with the learnable model parameters ~\citep{yang2018admm}. Ye et al. introduced a model-based neural network for estimating NODDI parameters ~\citep{ye2017tissue}, and we recently proposed a model-driven sparsity coding deep neural network (SCDNN) to estimate the IVIM parameters in the fetal brain ~\citep{TianshuZheng2021}.

However, convolution-based networks used in the current model-driven frameworks have a fixed perceptive field within a single layer ~\citep{luo2016understanding}, and repeatedly stacking deeper convolution layers will make the model bloated with sharply increasing computation load ~\citep{wang2018non}. Thus, the self-attention mechanism that adapts a dynamic perceptive field ~\citep{vaswani2017attention} can be added to the q-space deep learning task to improve its performance, which forms a Transformer-like structure. Because of its superior performance and flexibility, Transformer has gained immense interest in many fields. In image processing area, Vision Transformer (ViT) ~\citep{dosovitskiy2020image} has been introduced for classification tasks for computer vision and outperformed convolution networks. 

Despite its superior performance, applications of ViT in medical imaging are limited due to its high demand for training data. A typical ViT does not need any inductive bias ~\citep{dosovitskiy2020image} but requires a large quantity of data for training ($\sim$ 300M). The inductive bias can be considered as a priori hypothesis ~\citep{battaglia2018relational} that facilitates the network training process. In standard deep learning networks, convolution has the inductive bias of the locality and invariance of spatial translation. Recurrence has the inductive bias of the sequentiality and invariance of time translation. Graph network has the inductive bias of the arbitrarity with the invariance of the node or edge permutations. Inductive bias is not limited to these forms but can also be incorporated by tailored interventions into a deep neural network architecture ~\citep{karniadakis2021physics}, which can be introduced in the model-driven approach. 

In order to address the need for large amounts of data for training Transformer and to enhance the model interpretability, we add a new type of inductive bias—model bias, into the Transformer structure to drive the training process. In addition, sparse-coding is introduced to the model-driven process by converting the nonlinear dMRI models into a linear layout using a dictionary technique. Here, we propose a \emph{Microstructure Estimation Transformer with Sparse Coding} (METSC) for dMRI-based microstructural parameter estimation. This framework can be used to estimate different types of dMRI-based models, by modifying the decoder for a specific model. The major contributions are:

\begin{enumerate}
\item [1)] A new framework with the Transformer structure is proposed for dMRI model estimation, which is the first application of Transformer in a regression task in medical imaging.  
\item [2)] The METSC framework introduces a model bias, via the iterative structure, to allow Transformer to be trained with less data, e.g., only about 300K data.
\item [3)] This framework enables dMRI model parameter estimation using reduced q-space samples, and achieved a reduction of scan time up to 11.25 folds.
\item [4)] The proposed network has the superior performance in estimating the microstructure parameters, and reduced up to 81\% mean square error (MSE) compared with the former learning based method. 
\item [5)] We performed a thorough investigation of network architecture on two types of dMRI models with different data sizes, and METSC outperformed the other state-of-the-art q-space learning methods for both models.
\end{enumerate}

Specifically, we tested this new framework on the IVIM model, which is a bi-exponential model that separates the microcirculation in the capillaries from water diffusion in the tissues ~\citep{le2019can} using multiple b-value information, and the NODDI model ~\citep{zhang2012noddi} that estimates the microstructural dendrites and axons using multiple diffusion directions. Particularly, we investigated IVIM of the placenta, which is commonly used to access the placenta perfusion but is subject to abdominal and fetal motion; for NODDI, we used brain MRI from the \emph{Human Connectome Project} (HCP) ~\citep{van2013wu}. 

\section{Method}
\begin{figure*}[!htb] 
\centering 

\includegraphics[width=0.85\textwidth]{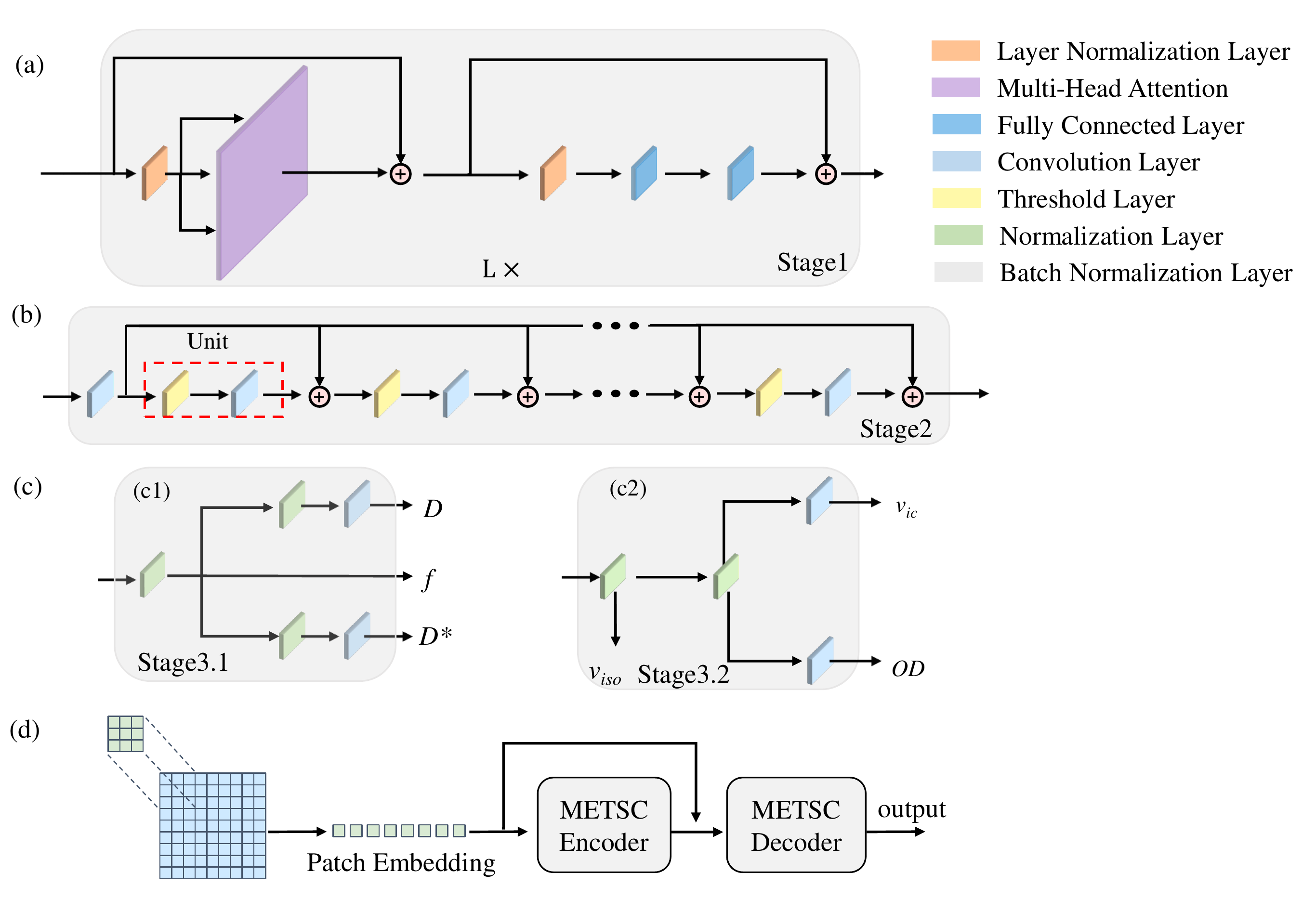} 

\caption{Overview of the METSC model. (a) Transformer-based encoder architecture, in which images are first split into patches and then fed into a ViT-like Transformer encoder, and additional skip connections are added to patch embedding and METSC encoder. (b) Schematic of the sparsity-based METSC decoder with a cascaded structure that includes model prior in training. A threshold layer plus a convolution layer make up the basic unit in the red dashed box. (c) Stages 3.1 and 3.2 show the maping stage designed for the IVIM and NODDI model outputs, respectively. (d) The entire METSC framework consists of the Transformer-based encoder and sparsity-based decoder.} %

\smallskip

\label{Figure1} 
\end{figure*}
In this section, we will first describe the IVIM model and the NODDI model, and then describe the METSC framework in detail.
\subsection{Background}
The rationale of selecting the IVIM model and the NODDI model is that IVIM requires multiple b-values densely sampled in the low b-value regime (b\textless800 $s/mm^{2}$), while NODDI relies on high-angular resolution (densely sampled diffusion directions) at high b-values (b\textgreater1000 $s/mm^{2}$). They are two representative models to test the generalizability of the proposed framework.
\subsubsection{IVIM}
The IVIM model separates the water diffusion in the tissue and pseudo-diffusion of microstructural flows in the capillaries, based on the different diffusivity of the two compartments ~\citep{le1988separation}, with a bi-exponential formulation:

\begin{equation}
    {{S}_{b}={S}_{0}\left[(1-f) e^{-b D}+f e^{-b D^{*}}\right]}  \label{eqIVIM}
\end{equation}

where $S_{0}$ is the non-diffusion-weighted signal and $S_{b}$ is the diffusion-weighted signals at a b-value of $b$; $f$ and $D^{*}$ are the fraction and pseudo-diffusivity of microcirculation, and $D$ is the water diffusivity in the tissues. 

Traditionally, the IVIM model can be fitted in two ways, including the two-step nonlinear least squares (NLLS) method ~\citep{federau2014measuring} and Bayesian method ~\citep{neil1993use}, and the latter considered to be the best method abdominal imaging~\citep{gustafsson2018impact}. Therefore, this study used the Bayesian fitting results of the fully sampled IVIM data as the gold standard, and compared the METSC with NLLS, Bayesian, and learning-based methods using downsampled IVIM data.
\subsubsection{NODDI}
The NODDI model separates the dMRI signal into three parts: intracellular, extracellular, and CSF compartments, and outputs microstructural parameters including the orientation dispersion (OD), and the volume fractions of the intra-cellular compartment ($v_{ic}$) and the CSF compartment ($v_{iso}$). The model can be written as follows:
\begin{equation}
    A=\left(1-v_{iso}\right)\left(v_{ic} A_{ic}+\left(1-v_{ic}\right) A_{e c}\right)+v_{iso} A_{iso} \label{eqNODDI}
\end{equation}
where $A$ is the normalized diffusion signal defined as $A= A_{b}/A_{0}$, with $A_{b}$ being the diffusion-weighted signal and $A_{0}$ being the non-diffusion-weighted signal. $A_{ic}$, $A_{ec}$ and $A_{iso}$ represent the signal contributions from the intra-cellular, extra-cellular, and CSF compartments, which can be defined as:

\begin{equation}
    \mathrm{~A}_{\mathrm{ic}}=\int_{S^{2}} \mathrm{M}\left(\frac{1}{2}, \frac{3}{2}, \kappa\right)^{-1} \mathrm{e}^{\kappa(\mu \cdot \mathbf{n})^{2}} \mathrm{e}^{-\mathrm{bd}_{\|}(\mathrm{q} \cdot \boldsymbol{n})^{2}} \mathrm{~d} \boldsymbol{n} 
\end{equation}
$A_{ic}$ is determined by the confluent hypergeometric function $M$, the diffusion encoding scheme \textbf{q} according to the gradient direction and b-value, concentration parameter $\kappa$, the mean orientation $\mu$, parallel diffusivity $\mathrm{d}_{\|}$, and $\mathrm{M}\left(\frac{1}{2}, \frac{3}{2}, \kappa\right)^{-1} e^{\kappa(\boldsymbol{\mu} \cdot \boldsymbol{n})^{2}} \mathrm{d} \boldsymbol{n}$  gives the probability of finding sticks along orientation $\boldsymbol{n}$. Beside \textbf{q}, $M$, ${{\kappa}}$, $\boldsymbol{\mu}$ , $\boldsymbol{n}$, $A_{ec}$ is also determined by $D\left ( \boldsymbol{n}\right) $, a cylindrically symmetric tensor with principal orientation $\boldsymbol{n}$. 
\begin{equation}
    \mathrm{A}_{\mathrm{ec}}=\exp \left(-\mathrm{b} \mathbf{q}^{\mathrm{T}}\left(\int_{S^{2}} \mathrm{M}\left(\frac{1}{2}, \frac{3}{2}, \kappa\right)^{-1} \mathrm{e}^{\kappa(\mu \cdot \boldsymbol{n})^{2}} \mathrm{D}(\boldsymbol{n}) \mathrm{d} \boldsymbol{n}\right) \mathbf{q}\right) 
\end{equation}
The CSF compartment is modeled as a Gaussian diffusion with a diffusivity of $d_{iso}$:
\begin{equation}
    \mathrm{A}_{iso} = \exp \left({-\mathrm{bd_{iso}}}\right)
\end{equation}
The original NODDI fitting with NLLS was relatively accurate but extremely slow. Daducci et al developed the Accelerated Microstructure Imaging via Convex Optimization (AMICO) toolkit ~\citep{daducci2015accelerated} that effectively speeded up the process via the sparse representation. Here we used NLLS fitted results of the fully sampled NODDI data as the gold standard, and compared the METSC with AMICO and learning-based methods using downsampled NODDI data.

\subsection{METSC}
The METSC framework (Fig. \ref{Figure1}) can be divided into three parts, a Transformer-based encoder and a sparsity-based decoder that consists of a sparse coding neural network and a model-specific microstructural mapping network.

\subsubsection{Transformer based METSC encoder}
The Transformer-based METSC encoder is adapted from the ViT structure ~\citep{dosovitskiy2020image} and tailored for the model estimation task. It consists of two layer-normalization layers ~\citep{ba2016layer}, a multi-head attention layer, and two fully connected layers (Fig. \ref{Figure1}(a)). To accelerate the training of the METSC encoder, a skip connection is added ~\citep{he2016deep} to connect the beginning of one encoder to the end of the encoder. Also to adapt the classification task to the estimation task for dMRI models, the BERT’s [class] token is removed. Following the selection of nonlinear activation function in BERT ~\citep{devlin2018bert}, GELU is chosen as our activation function. 

First of all, similar to ViT, to accommodate the 2D input, the image $ \mathrm{q}_{0} \in \mathbb{R}^{H \times W \times C} $ (where $H$ and $W$ are the height and width of the image, $C$ is the number of channels which is the number of b-values in our network) is split into smaller 2D patches $ \mathrm{q}_{p} \in \mathbb{R}^{N \times{H}_{p} \times {W}_{p} \times C} $ (where  $ {H}_{p} $ and $ {W}_{p} $  are the height and width of the patch,  $N = {H \mathord{\left/ {\vphantom {H {{H_P} \times }}} \right. \kern-\nulldelimiterspace} {{H_P} \times }}{W \mathord{\left/  {\vphantom {W {{W_P}}}} \right.  \kern-\nulldelimiterspace} {{W_P}}}\ $ is the number of patches) with a non-overlap design, as the sequence of input to the Transformer. Patch embedding is applied to the sequence of patches, which is learned through training with a linear projection in a fully connected layer. 
\begin{equation}
    \boldsymbol{\mathrm{z}_{0}}=\left[\mathrm{q}_{p}^{1} \mathbf{E} ; \mathrm{q}_{p}^{2} \mathbf{E} \cdots \mathrm{q}_{p}^{N} \mathbf{E}\right]
\end{equation}
where {$\boldsymbol{z_{0}}$} is the sequence of embedded patches and $ \mathrm{E} \in \mathbb{R}^{({H}_{p} \times {W}_{p} \times C) \times D} $ is the patch embedding projection matrix. Then the data are sent to a Layer Normalization layer followed by the multi-head self-attention (MSA).
\begin{equation}
    \boldsymbol{z_{M}}=\mathrm{MSA}\left(\boldsymbol{z_{l}}\right)=\left[\mathrm{SA}_{1}\left(\boldsymbol{z_{l}}\right), \mathrm{SA}_{2}\left(\boldsymbol{z_{l}}\right) \cdots \mathrm{SA}_{k}\left(\boldsymbol{z_{l}}\right)\right] \mathbf{U}_{m s a} 
\end{equation}
where, $\boldsymbol{z_{l}}$ is the normalized $\boldsymbol{z_{0}}$, $SA\left ( \cdot  \right )$ is the self-attention layer, $\mathbf{U}_{msa} \in \mathbb{R}^{k \times {D}_{h} \times D} $ with k being the number of heads, $D$ being the dimension of the fully connected layer, and $D_{h}$ being the scaler that is typically set to $D/k$.
\begin{equation}
    \mathrm{SA}\left(\boldsymbol{z_{l}}\right)=\operatorname{softmax}\left(q k^{T} / \sqrt{D_{h}}\right)v
\end{equation}
where $q$, $k$, $v$ correspond to the query, key, and value of the input sequence $z_{l}$ and they can be calculated as follows:
\begin{equation}
    \left[q, k, v\right] = \boldsymbol{z_{l}} \boldsymbol{\mathrm{U}}_{qkv},   \qquad \boldsymbol{\mathrm{U}}_{qkv} \in \mathbb{R}^{{D} \times {3D}_{h}} 
\end{equation}
The input of embedded patches and the output from the MSA are connected through the skip connection. The output of the skip connection is sent into the Layer Normalization layer and two fully connected layers with the gaussian error linear units (GELU) activation function and dropout. \begin{equation}
    \boldsymbol{z}_{FC} = \mathrm{FFN} \left( \boldsymbol{z}_{l2} \right) + \boldsymbol{z}_{0} + \boldsymbol{z}_{M} 
\end{equation}

Where, $\boldsymbol{z}_{l2}$ is the normalized $\boldsymbol{z}_{0} + \boldsymbol{z}_{M} $, and $\mathrm{FFN \left( \cdot \right)}$ is a feedforward network containing two fully connected layers. The flow chart of the Transformer-based METSC encoder is shown in Fig. \ref{Figure2}(a).
The overall procedure is summarized in the supplementary material Algorithm 1. 

\begin{figure}[!htb]
\centering

\includegraphics[width=0.47\textwidth]{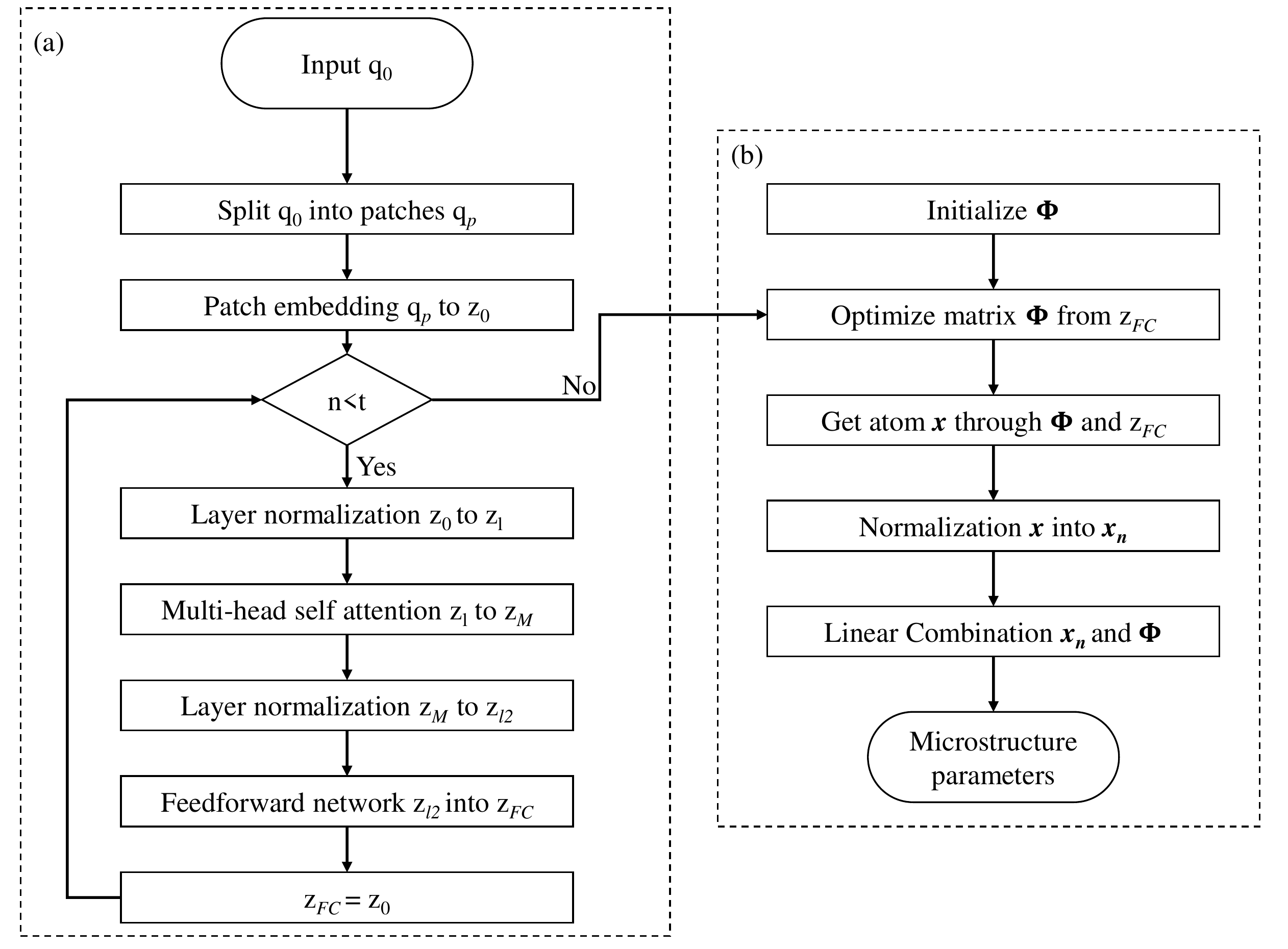} 

\caption{The algorithmic flow of the METSC. (a) Transformer-based encoder flow chart. (b) Sparsity-based decoder flow chart. Here t is the number of Transformer-based encoders.} %

\smallskip

\label{Figure2} 
\end{figure}
\subsubsection{Sparsity-based METSC decoder}
The sparsity-based METSC decoder is a model-driven deep neural network that provides an inductive bias in METSC. As mentioned above, the inductive bias can be seen as a type of prior information ~\citep{battaglia2018relational,karimi2021deep}, which takes the form of the IVIM model or NODDI model in the current study. In this section, the METSC decoder configuration will be shown for IVIM model and NODDI model, separately. The method of construction sparsity-based METSC decoder can be briefly summarized as dictionary construction and linear combination.
\paragraph{Decoder for IVIM Model} \
\label{Decoder IVIM}

\textbf{Sparse coding}. The main challenge of a model-driven network is the choice of the optimization algorithm. The IVIM model follows a nonlinear bi-exponential function (Eq. [\ref{eqIVIM}]), in which the $D$ and $D^{*}$ are the exponential terms, and $f$ is coupled with exponents. Thus, the model cannot directly be translated into the network through the optimization procedure. Inspired by AMICO ~\citep{daducci2015accelerated}, the nonlinear models can be represented via dictionary learning. In this work, the IVIM model is linearized through a sparse-coding based dictionary learning framework as below:
\begin{equation} 
\boldsymbol{z}_{FC}=\boldsymbol{\mathrm{\Phi} x}   + \boldsymbol{\eta} \label{eq11}
\end{equation} 
where $\boldsymbol{z}_{FC}=\left(z_{1}, \cdots, z_{n}\right)^T $ is a vector comprised of the encoded dMRI signals from the Transformer-based encoder that is acquired at n different b-values; $\boldsymbol{\Phi}$ is a dictionary vector ($\boldsymbol{\Phi} \in \mathbb{R}^{1 \times 2j}$, where $j$ corresponds to the length of the discretized $D$ and $D^{*}$); $\boldsymbol{x}$ is a vector of the dictionary coefficients ($\boldsymbol{x} \in \mathbb{R}^{2j \times 1}$), and $\boldsymbol{\eta}$ is a noise term. The dictionary can be established through the discretized $D$ and $D^{*}$, and the $x$ corresponds to the fraction of $f$:
\begin{equation} 
\boldsymbol{\Phi}=\left[\boldsymbol{\Phi_{D}}, \boldsymbol{\Phi_{D^{*}}}\right] 
\end{equation} 
\begin{equation} 
\boldsymbol{x}=\left[\boldsymbol{x_{1-f}}, \boldsymbol{x_{f}}\right]^{T} 
\end{equation} 
The signals are normalized to the b0 signal and fall into the interval of [0,1], and thus, the three parameters can be reformulated as below:
\begin{equation}
\boldsymbol{{x}}=\frac{\boldsymbol{x}+\tau}{\|\boldsymbol{x}+\tau\|_{1}} \label{xnormal} 
\end{equation}
\begin{equation}
\boldsymbol{{x}}_{1-f}=\frac{\boldsymbol{{x}}_{1-f}+\tau}{\|\boldsymbol{{x}}_{1-f}+\tau\|_{1}} \label{x1-f}  
\end{equation} 
\begin{equation}
\boldsymbol{{x}}_{f}=\frac{\boldsymbol{{x}}_{f}+\tau}{\|\boldsymbol{{x}}_{f}+\tau\|_{1}} \label{xf}
\end{equation}
\begin{equation}
f=\boldsymbol{\mathrm{I}_{1}x} \label{eq17}
\end{equation}
\begin{equation}
D=\frac{\boldsymbol{\Phi} \boldsymbol{\mathrm{I_{2}}} \boldsymbol{x}_{1-f}}{\boldsymbol{\mathrm{I_{2}}} \boldsymbol{x}_{1-f}} \label{eq18}
\end{equation}
\begin{equation}
D^{*}=\frac{\boldsymbol{\Phi \mathrm{I}_{3}} \boldsymbol{x}_{f}}{\boldsymbol{\mathrm{I}}_{1} \boldsymbol{x}_{f}}  \label{eq19}
\end{equation}
where, $\tau=1e^{-10}$ is set to avoid 0 in the denominator. $\boldsymbol{\mathrm{I_{1}}} \in \mathbb{R}^{1 \times 2j}$, $\boldsymbol{\mathrm{I_{2}}}$, $\boldsymbol{\mathrm{I_{3}}} \in \mathbb{R}^{2j \times j}$, and $\boldsymbol{\mathrm{I_{4}}} \in \mathbb{R}^{1 \times 2j}$ are defined as:
\begin{equation}
    \boldsymbol{\mathrm{I_{4}}}=\left[0 \cdots 0,1 \cdots 1\right]^{T}
\end{equation}
\begin{equation}
    \boldsymbol{\mathrm{I_{2}}}= \begin{pmatrix} \boldsymbol{\mathrm{I_{5}}} \\ \boldsymbol{0} \end{pmatrix}
\end{equation}
\begin{equation}
    \boldsymbol{\mathrm{I_{3}}}= \begin{pmatrix}  \boldsymbol{\mathrm{0}} \\\boldsymbol{\mathrm{I_{5}}} \end{pmatrix}
\end{equation}
\begin{equation}
    \boldsymbol{\mathrm{I_{4}}}=\left[1 \cdots 1,0 \cdots 0\right]^{T}
\end{equation}
where, $\boldsymbol{\mathrm{I_{5}}} \in \mathbb{R}^{j \times j}$  is an identity matrix.

\textbf{Network construction.} According to Eq.[\ref{eq11}], the bi-exponential IVIM model can be converted into a linear model. The next step is to establish a dictionary that optimally represents the signals, with the objective function as below:
\begin{equation}
    \boldsymbol{\min _{x}\left\|y-\Phi x\right\|_{2}^{2}}+\beta\boldsymbol{\|x\|_{0}}       
\end{equation}
where $\beta$ controls the sparsity of matrix $\boldsymbol{x}$. Here, the \emph{Iterative Hard Thresholding} (IHT) ~\citep{blumensath2009iterative} method is used for optimization, which is formulated:

\begin{equation}
    \boldsymbol{x}^{n+1}=H_{M}\left(\boldsymbol{\mathrm{W}}{\boldsymbol{z}_{FC}}+\boldsymbol{\mathrm{S}} \boldsymbol{x}^{n}\right) 
\end{equation}

\begin{equation}
 = H_{M}\left[\boldsymbol{x}^{n}+\boldsymbol{\Phi}^{H}(\boldsymbol{z}_{FC}-\boldsymbol{\Phi}) \boldsymbol{x}^{n}\right]
\end{equation}

\begin{equation}
    = H_{M}\left[\boldsymbol{\Phi}^{H} \boldsymbol{z}+\left( \boldsymbol{\mathrm{I}}-\boldsymbol{\Phi}^{H} \boldsymbol{\Phi}\right) \boldsymbol{x}^{n}\right]
\end{equation}
where, $\boldsymbol {\mathrm{W}={\Phi}^{H}}$, $\boldsymbol {\boldsymbol{\mathrm{S}}=\mathrm{I}-{\Phi}^{H}\Phi}$, and $H_M$ denotes a nonlinear operator:
\begin{equation}
   H_{M}\left(\boldsymbol{x}\right)\left\{\begin{array}{ll}
0 & \text { if }\left|\boldsymbol{x}\right|<\lambda \\
\boldsymbol{x} & \text { if }\left|\boldsymbol{x}\right| \geq \lambda
\end{array}\right.                               \label{opertaor1}
\end{equation}
where {$\lambda$} is a positive threshold. In the IVIM model, the model parameters are nonnegative, and thus, the nonlinear operator can be simplified as Eq.[\ref{eq29}]:
\begin{equation}
    H_{M}\left(\boldsymbol{x}\right)\left\{\begin{array}{ll}
0 & \text { if }\boldsymbol{x}<\lambda \\
\boldsymbol{x} & \text { if }\boldsymbol{x} \geq \lambda
\end{array}\right.     \label{eq29}                          
\end{equation}
Thus, the network can be designed by unfolding the iterative process using the sparsity-based METSC decoder. $\boldsymbol{\mathrm{W}}$ and $\boldsymbol{\mathrm{S}}$ are the shared weights among the layers including the dictionary layer, which do not need to be pre-trained. The decoder design algorithm for IVIM can be summarized in the supplementary material Algorithm 2.1. After the dictionary is trained in Fig. \ref{Figure1}(b), the parameters can be estimated through Eq. [\ref{eq17}], Eq. [\ref{eq18}], and Eq. [\ref{eq19}], as shown in Fig. \ref{Figure1}(c1). 

\paragraph{Decoder for NODDI Model} \

\textbf{Sparse coding.} In the NODDI model (Eq.[\ref{eqNODDI}]), the signal can also be linearized following Eq. [\ref{eq11}].
Similarly in NODDI, $\boldsymbol{z}_{FC}=\left(z_{1}, \cdots, z_{n}\right)^T $ is a vector comprised of the encoded dMRI signals from the Transformer-based encoder that is acquired at n different diffusion gradients; $\boldsymbol{\Phi}$ is a dictionary vector ($\boldsymbol{\Phi} \in \mathbb{R}^{1 \times 2j+i}, \ \boldsymbol{\Phi_{t}} \in \mathbb{R}^{1 \times 2j},\ \boldsymbol{\Phi}_{i} \in \mathbb{R}^{1 \times i} $ here $j$ corresponds to the length of the discretized $v_{ic}$ and $\kappa$, and the length of $v_{iso}$ is $i$), and $\boldsymbol{x}$ is a vector of the dictionary coefficients ($\boldsymbol{x} \in \mathbb{R}^{1 \times 2j+i},\ \boldsymbol{x}_{t} \in \mathbb{R}^{2j \times 1},\ \boldsymbol{x}_{i} \in \mathbb{R}^{i \times 1}$, $\boldsymbol{x}_{t}$  is the coefficient of anisotropic signals including $v_{ic}$, $\kappa$, and $\boldsymbol{x}_{i}$ is the coefficient of the isotropic $v_{iso}$. The dictionary can be established through the discretized $v_{ic}$, $\kappa$, $v_{iso}$, and $\boldsymbol{\Phi}$, $\boldsymbol{x}$ can be defined below:
\begin{equation}  
\boldsymbol{\Phi}=\left[\boldsymbol{\Phi}_{t}, \boldsymbol{\Phi}_{i}\right]  
\end{equation} 
\begin{equation}  
\boldsymbol{x}=\left[\boldsymbol{x}_{t}, \boldsymbol{x}_{i}\right]^{T}  
\end{equation} 
The components in $\boldsymbol{x}_{t}$ need to be normalized into the interval of [0,1]. Then, the three parameters $v_{ic}$, $\kappa$ and $v_{iso}$ can be obtained as below:
\begin{equation}
    v_{iso}=\mathbf{I}_{6} \boldsymbol{x}_{i} \label{eq32}
\end{equation}
\begin{equation}
    \boldsymbol{x}_{t}=\frac{\boldsymbol{x}_{t}+\tau}{\left\|\boldsymbol{x}_{t}+\tau\right\|_{1}}
\end{equation}
\begin{equation}
v_{ic}=\frac{\boldsymbol{\Phi}_{t} \mathbf{I}_{7} \boldsymbol{x}_{t}}{\mathbf{I}_{9} \boldsymbol{x}_{t}}  \label{eq34}
\end{equation}
\begin{equation}
\kappa=\frac{\boldsymbol{\Phi}_{t} \mathbf{I}_{8} \boldsymbol{x}_{t}}{\mathbf{I}_{9} \boldsymbol{x}_{t}}  
\end{equation}
where $\mathbf{I}_{5} \in \mathbb{R}^{j \times j}$ is an identity matrix, and $\mathbf{I}_{6} \in \mathbb{R}^{1 \times i}, \mathbf{I}_{7} \in \mathbb{R}^{2 j \times 2 j}, \mathbf{I}_{8} \in \mathbb{R}^{2 j \times 2 j}, \mathbf{I}_{9} \in \mathbb{R}^{1 \times 2 j}$ are defined as:
\begin{equation}
  \mathbf{I}_{6}=\left[ 1 \cdots  1 \right]^{T}  
\end{equation}
\begin{equation}
\mathbf{I}_{7}=\left(\begin{array}{cc}
\mathbf{I}_{5} & 0 \\
0 & 0
\end{array}\right)    
\end{equation}
\begin{equation}
  \mathbf{I}_{8}=\left[ 1 \cdots  1 \right]^{T}  
\end{equation}
\begin{equation}
\mathbf{I}_{9}=\left(\begin{array}{ll}
0 & 0 \\
0 & \mathbf{I}_{5}
\end{array}\right)
\end{equation}
Finally, OD can be calculated through:
\begin{equation}
    \mathrm{OD}=\frac{2}{\pi} \operatorname{atctan}\left(\frac{1}{\kappa}\right) \label{eq40}
\end{equation}

\textbf{Network construction.} The network can be designed in the same way as in Section \ref{Decoder IVIM}. After the dictionary is trained in Fig. \ref{Figure1}(b), the parameters can be estimated through Eq.[\ref{eq32}], Eq.[\ref{eq34}], and Eq.[\ref{eq40}] as shown in Fig. \ref{Figure1}(c2). The decoder design algorithm for NODDI can be summarized in the supplementary material Algorithm 2.2. Overall, the flow chart of the METSC decoder can be shown in Fig. \ref{Figure2}(b). 
Finally, the METSC encoder and decoder can be combined with a skip connection to connect the input and the framework, as illustrated in Fig. \ref{Figure1}(d). The code will be provided at https://github.com/Tianshu996/METSC upon publication.
\section{Results}
\begin{table*}[htbp]
\caption{ MSE of estimated $f$, $D$, and $D^{*}$ using METSC with different encoders, decoders, and input forms, on 9 testing data (about 248059 voxels). *p\textless 0.05, **p\textless 0.01, ***p\textless 0.001 by paired t-test between different methods on the nine patients.}
\centering
\begin{tabular}{ccccc}
\hline
\multirow{2}{*}{} &
  \multirow{2}{*}{} &
  \multirow{2}{*}{$f \times 10^{-6}$} &
  \multirow{2}{*}{$D  (\times 10^{-4} {\mu m}^{2}/ms)$ } &
  \multirow{2}{*}{$D^{*}  (\times 10^{-2} {\mu m}^{2}/ms)$} \\
 &              &      &       &     \\ \hline
\multirow{4}{*}{\begin{tabular}[c]{@{}c@{}}Decoder\\ (with METSC encoder /  \\ patched input)\end{tabular}} &
  \multirow{2}{*}{Model-free} &
  \multirow{2}{*}{46.5(**)} &
  \multirow{2}{*}{4.59 (*)} &
  \multirow{2}{*}{47(**)} \\
 &              &      &       &     \\
  &              &      &       &     \\

 & METSC        & 6.3  & 2.15  & 1.4 \\ \hline

\multirow{4}{*}{\begin{tabular}[c]{@{}c@{}}Encoder\\ (with METSC decoder / \\ patched input)\end{tabular}} &
  \multirow{2}{*}{Convolutional} &
  \multirow{2}{*}{8.45 (**)} &
  \multirow{2}{*}{2.78} &
  \multirow{2}{*}{8.1 (**)} \\
 &              &      &       &     \\
   &              &      &       &     \\
 & METSC        & 6.33 & 2.15  & 1.4 \\ \hline
\multirow{4}{*}{\begin{tabular}[c]{@{}c@{}}Input\\ (with METSC encoder / \\ METSC decoder )\end{tabular}} &
  \multirow{2}{*}{Non-patch} &
  \multirow{2}{*}{$3.60 \times 10^3$ (***)} &
  \multirow{2}{*}{$2.44 \times 10^6$ (***)} &
  \multirow{2}{*}{$1.99 \times 10^3$ (***)} \\
 &              &      &       &     \\
   &              &      &       &     \\
 & Patch        & 6.33 & 2.151 & 1.4 \\ \hline

\end{tabular} \label{Tab1}
\end{table*}
In this section, we tested the METSC framework on both the IVIM and NODDI models to find the optimal setup. We also compared our method with the state-of-the-art networks for dMRI-based microstructural estimation.
\subsection{IVIM Model}
In subsection, we used placental IVIM data to train the model and determined the optimal hyper-parameter with a set of ablation experiments. We then compared our approach to existing optimization-based and learning-based methods. We also evaluate the generalizability of our framework on an independent dataset obtained from a different center. 
\subsubsection{Dataset}
The placental IVIM data were acquired on a 1.5T GE scanner (SIGNA HDXT) from 24 normal pregnant women (gestational age 13-37 weeks) under Institional Research Board approval at the local hospital. Diffusion gradients were applied in three orthogonal directions at 10 b-values of $0, 10, 20, 50, 80, 100, 150, 200, 300, 500$ $s/{mm}^{2}$ with the following acquisition parameters: repetition time/echo time = $3000/76$ $ms$, in-plane resolution =  ${1.25} \times {1.25}$ ${mm}^{2}$, field of view = $ {320} \times {320}$ ${mm}^{2}$, 15 slices with a slice thickness of $ 4 mm$. The data was preprocessed through bias correction and registration for motion correction, and signals within the placenta mask were used in the following experiment in a voxelwise manner. To obtain the gold standard ($f$, $D$, and $D^{*}$), we performed Bayesian estimation of these parameters using the full dataset (10 b-values) using the Matlab fitting toolbox ~\citep{gustafsson2018impact}.

The dataset was then divided into the training, validation, and testing datasets, resulting in 324016 voxels from 15 subjects for training with 10\% of the training samples used for validation ~\citep{golkov2016q}, and 248059 voxels from 9 subjects for testing. All datasets were split into overlapping patches with a step size of 1 in zero-padded images.

We also generated simulation data (S) using the gold standard model parameters according to Eq.[\ref{eq41}] and added noise to generate data at different signal-to-noise ratio (SNR) levels according to ~\citep{daducci2014sparse}:
\begin{equation}
    {{S}_{\text {simulated }}=\sqrt{\left({S}+\xi_{1}\right)^{2}+\left(\xi_{2}\right)^{2}}}   \label{eq41} 
\end{equation}
where, $\xi_{1}, \xi_{2} \sim N\left(0, \sigma^{2}\right)$, and $\sigma=S_{0}/ \mathrm{SNR}$. Similar to  ~\citep{daducci2014sparse}, we assumed $S_{0}=1$ and SNR varied from 10 to 70.
\subsubsection{Training}
All the models were trained using Adam as the optimizer with the total epochs of 2000 and batch size of 512. We used the cosine warm-up method in the first 200 epochs and a reducing learning rate with an initial learning rate of $1\times 10^{-4}$. The experiments were performed on a Linux machine with eight NVIDIA GeForce RTX 3090 GPUs.
\subsubsection{Ablation Experiments on Network Architecture}
\label{ablation experiments}
Eight pairs of ablation experiments were performed to compare model-free decoder / METSC decoder, convolution encoder / METSC encoder, non-patched / patched image inputs, different sizes of the training data, different combinations of b-value, the different number of b-values, the different dictionary size, and the different patch size.

\begin{figure*}[!htb] 
\centering 

\includegraphics[width=0.8\textwidth]{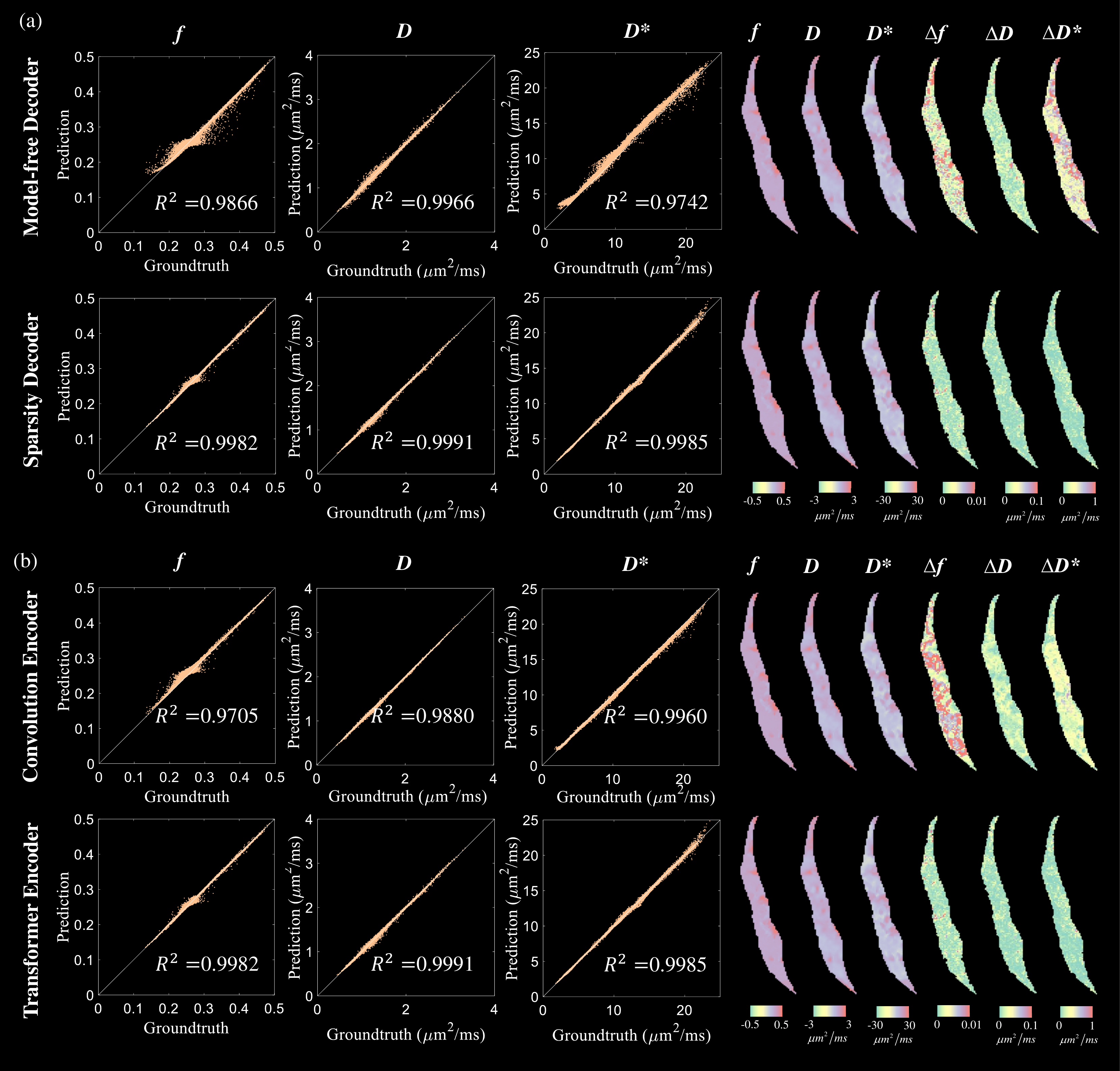} 

\caption{Ablation experiments on the performance of decoder and encoder of the METSC framework. (a) Estimated model parameters using the model-free versus sparsity-based METSC decoders, based on the voxelwise correlation between the estimated values and ground truth, the estimated parameter maps ($f$, $D$, and $D^{*}$), and the error maps ($\Delta f$, $\Delta D$, and $\Delta D^{*}$). (b) Estimated model parameters using the convolution and Transformer-based METSC encoders.} %
\smallskip

\label{Fig3} 
\end{figure*}

(1)\textbf{ Model-free decoder versus Sparsity decoder} were tested in combination with METSC encoder with patched inputs at five b-values ($20, 50, 150, 300, 500 s/mm^2$). The model-free decoder was designed following Golkov that was composed of three fully connected layers and the nonlinear activation ReLU ~\citep{golkov2016q}. The results in Fig. \ref{Fig3}(a) showed the model parameters estimated from the METSC decoder achieved higher correlations with the gold standard compared to the model-free decoder. It was also clear from the error maps (gold standard $-$ estimated parameters, Fig. \ref{Fig3}(a)) that the METSC decoder resulted in lower estimation errors than the model-free decoder, especially for the $\Delta f$ and the $\Delta D^{*}$ maps.

(2)\textbf{ Convolution encoder versus Transformer encoder} were tested in combination with METSC decoder with patched inputs at five b-values ($20, 50, 150, 300, 500 s/mm^2$). The convolution encoder consisted of 2D convolution layers, Batch Normalization layers, and the ReLU activation. The results in Fig. \ref{Figure2}(b) showed that the Transformer encoder provided a more accurate estimation of the IVIM model parameters according to the correlation plots and the $\Delta f$, $\Delta D$, and $\Delta D^{*}$ maps. 

(3)\textbf{ Patch versus non-patch-based inputs} were tested with the METSC encoder and decoder at five b-values ($20, 50, 150, 300, 500 s/mm^2$). The results in Table \ref{Tab1} (third row) showed that METSC with patch-based inputs outperformed the non-patch inputs. 

\begin{table}[]
\centering
\caption{The MSE of estimated $f$, $D$, and $D^{*}$ using the different number of training data with METSC and ViT.}
\resizebox{3.45in}{!}{%
\begin{tabular}{ccccccc}
\hline
\multirow{2}{*}{} & \multicolumn{2}{c}{\multirow{2}{*}{$f (\times 10^{-4})$}} & \multicolumn{2}{c}{\multirow{2}{*}{$D  (\times 10^{-4} {\mu m}^{2}/ms)$ }} & \multicolumn{2}{c}{\multirow{2}{*}{$D^{*}  (\times 10^{-2} {\mu m}^{2}/ms)$ }} \\
     & \multicolumn{2}{c}{} & \multicolumn{2}{c}{} & \multicolumn{2}{c}{} \\ \cline{2-7} 
     & ViT   & METSC   & ViT   & METSC    & ViT    & METSC   \\ \hline
10K  & 7.2        & 0.83    & $2.02\times 10^{6}$  & $2.02\times 10^{3}$  & 630   & 39      \\
40K  & 2.1        & 0.29    & 400       & 3.9      & 160    & 9.5     \\
200K & 2.3        & 0.11    & 4.34     & 2.2      & 171   & 2.6     \\
300K & 2.8        & 0.061   & 3.93     & 1.8      & 241    & 1.3     \\ \hline
\end{tabular}%
} \label{Tab2}
\end{table}

\begin{figure}[!htb] 
\centering 

\includegraphics[width=0.45\textwidth]{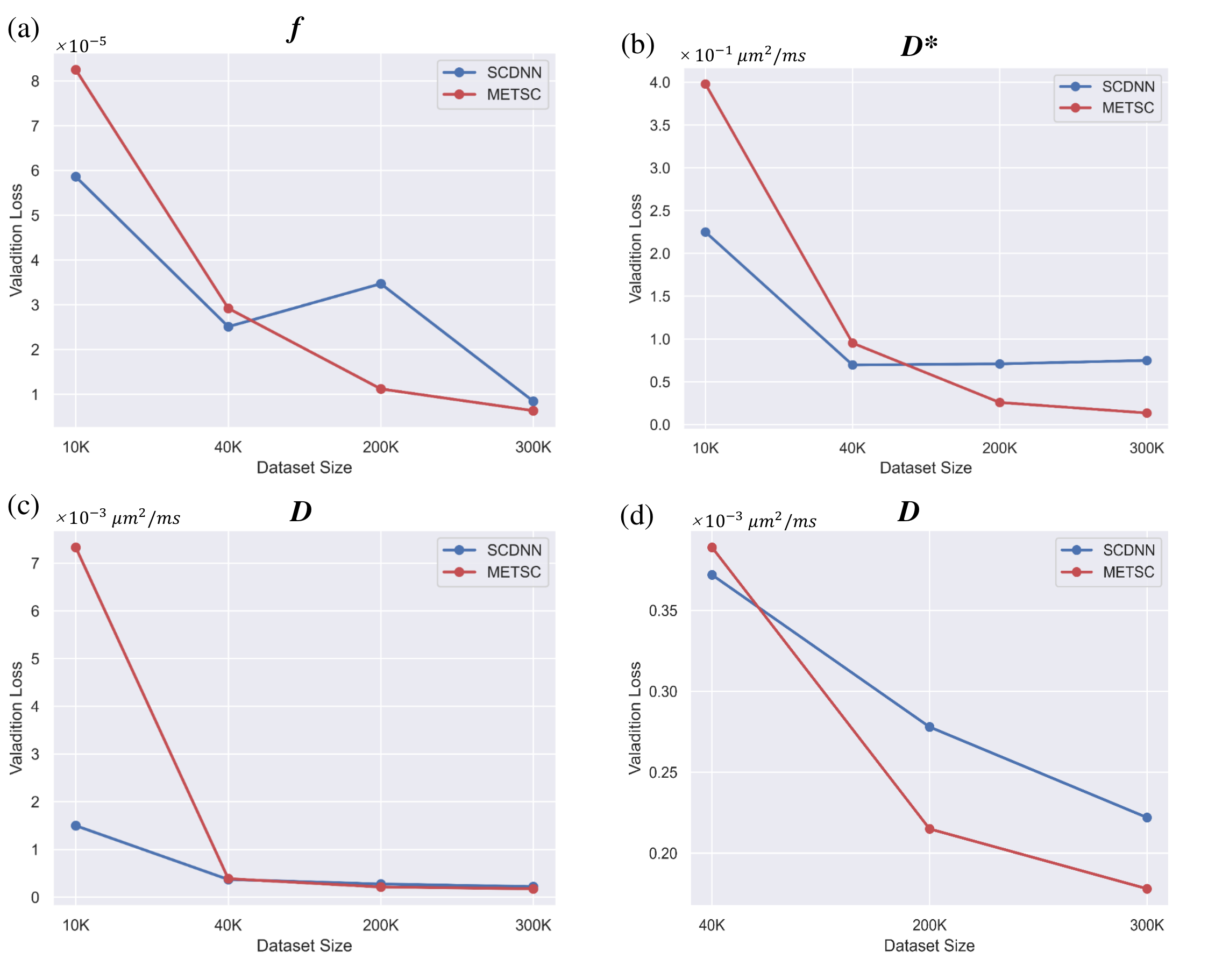} 

\caption{(a-c) The MSE of estimated $f$, $D$, and $D^{*}$ using the different number of training data with METSC and SCDNN, which is a model-driven learning method without the Transformer encoder~\citep{TianshuZheng2021}. (d) Zoom-in fiew of the MSE of $D$ in the range of training data size 40K-300K.} %

\smallskip

\label{Fig5} 
\end{figure}

(4)\textbf{ Effect of training data size.} Based on the optimal METSC setup (Transformer encoder + Sparsity decoder with patched inputs) obtained above, we tested the network performance using varying training data sizes of 10K, 40K, 200K, 300K. We found that with insufficient training data (10K), METSC performed worse than the SCDNN, which is a model-driven neural network without the Transformer encoder ~\citep{TianshuZheng2021}. As the training data increased to 200K, METSC reached a comparable accuracy to the SCDNN. The performance of METSC further increased and outperformed SCDNN as the number of training data increased to 300K (Fig. \ref{Fig5}). We also tested the ViT structure without sparsity decoder, and showed that METSC outperformed the ViT structure for all training data sizes between 10K-300K (Table \ref{Tab2}). 

\begin{table}[!htb]
\centering

\caption{The MSE of estimated IVIM model parameters using different combinations of b-values and different number of b-values. The combinations with top performance were highlighted in bold.}
{
\scalebox{0.7}{
\begin{tabular}{@{}cccc@{}}
\hline
                                         & $f  $     & $D  $    & $D^{*} $   \\ 
\multicolumn{1}{l}{} & \multicolumn{1}{l}{($\times 10^{-4})$} & \multicolumn{1}{l}{$(\times 10^{-4} {\mu m}^{2}/ms)$} & \multicolumn{1}{l}{$(\times 10^{-2} {\mu m}^{2}/ms)$} \\ \hline
\multirow{1}{*}{3   b-values}                  & \multirow{2}{*}{$0.41$}           & \multirow{2}{*}{$2.84$}                               & \multirow{2}{*}{$8.4$}                                 \\
  \multirow{1}{*}{(20, 150, 500)$s/mm^2$ }                                  &                                      &                                                          &                                                           \\
\multirow{1}{*}{5   b-values Comb1 (ours)}   & \multirow{2}{*}{\textbf{0.063}}  & \multirow{2}{*}{$2.2$}                               & \multirow{2}{*}{\textbf{1.4}}                        \\
 \multirow{1}{*}{(20, 50, 150, 300, 500)$s/mm^2$ }                                                              &                                      &                                                          &                                                           \\
\multirow{1}{*}{5   b-values Comb2}   & \multirow{2}{*}{$0.4$}           & \multirow{2}{*}{$2.1$}                               & \multirow{2}{*}{$1.8$}                                 \\
   \multirow{1}{*}{(20, 50, 150, 200, 500)$s/mm^2$ }                                                           &                                      &                                                          &                                                           \\
\multirow{1}{*}{5   b-values Comb3}   & \multirow{2}{*}{\textbf{0.063}}  & \multirow{2}{*}{$5.6$}                               & \multirow{2}{*}{$1.8$}                                 \\
     \multirow{1}{*}{(20, 50, 200, 300, 500)$s/mm^2$ }                                                         &                                      &                                                          &                                                           \\
\multirow{1}{*}{5   b-values Comb4}  & \multirow{2}{*}{$0.097$}           & \multirow{2}{*}{\textbf{1.7}}                      & \multirow{2}{*}{$1.7$}                                \\
        \multirow{1}{*}{(20, 100, 150, 300, 500)$s/mm^2$ }                                                      &                                      &                                                          &                                                           \\
\multirow{1}{*}{5   b-values Comb5}   & \multirow{2}{*}{$0.094$}           & \multirow{2}{*}{\textbf{1.7}}                      & \multirow{2}{*}{$2.0$}                                 \\
          \multirow{1}{*}{(20, 80, 150, 300, 500)$s/mm^2$ }                                                    &                                      &                                                          &                                                           \\
\multirow{1}{*}{7 b-values} & \multirow{2}{*}{$0.057$}            & \multirow{2}{*}{$1.1$}                               & \multirow{2}{*}{$1.9$}                                 \\
          \multirow{1}{*}{(20, 50, 100, 150, 200, 300, 500)$s/mm^2$ }                                                    &                                      &                                                          &                                                           \\ \hline
\end{tabular}}} \label{Tab3}
\end{table}

(5)\textbf{ Effect of the choice of b-values.} In Experiment 1-4, we used a selected subset of 5 (out of 10) b-values at $20, 50, 150, 300, 500 s/mm^2$ as the diffusion-weighted signals at these b-values best characterize the signal decay curve. Here we tested another four combinations of five b-values as listed in Table \ref{Tab3}, and the combinations with top performance were highlighted in bold for each model parameter. Overall, the b-value combination of $20, 50, 150, 300, 500 s/mm^2$ achieved the optimal balance of estimation accuracy for all the three IVIM parameters. 

(6)\textbf{ Effect of the number of b-values.} As expected, the estimation accuracy increased with the number of b-values. Table \ref{Tab3} showed that the MSE considerably reduced as the number of b-values increased from 3 to 5, but the further increase of b-values from 5 to 7 had a limited improvement (15\% reduction in the sum of validation loss) at the expense of 1.4 longer scan time. 

\begin{figure}[!htb] 
\centering 

\includegraphics[width=0.45\textwidth]{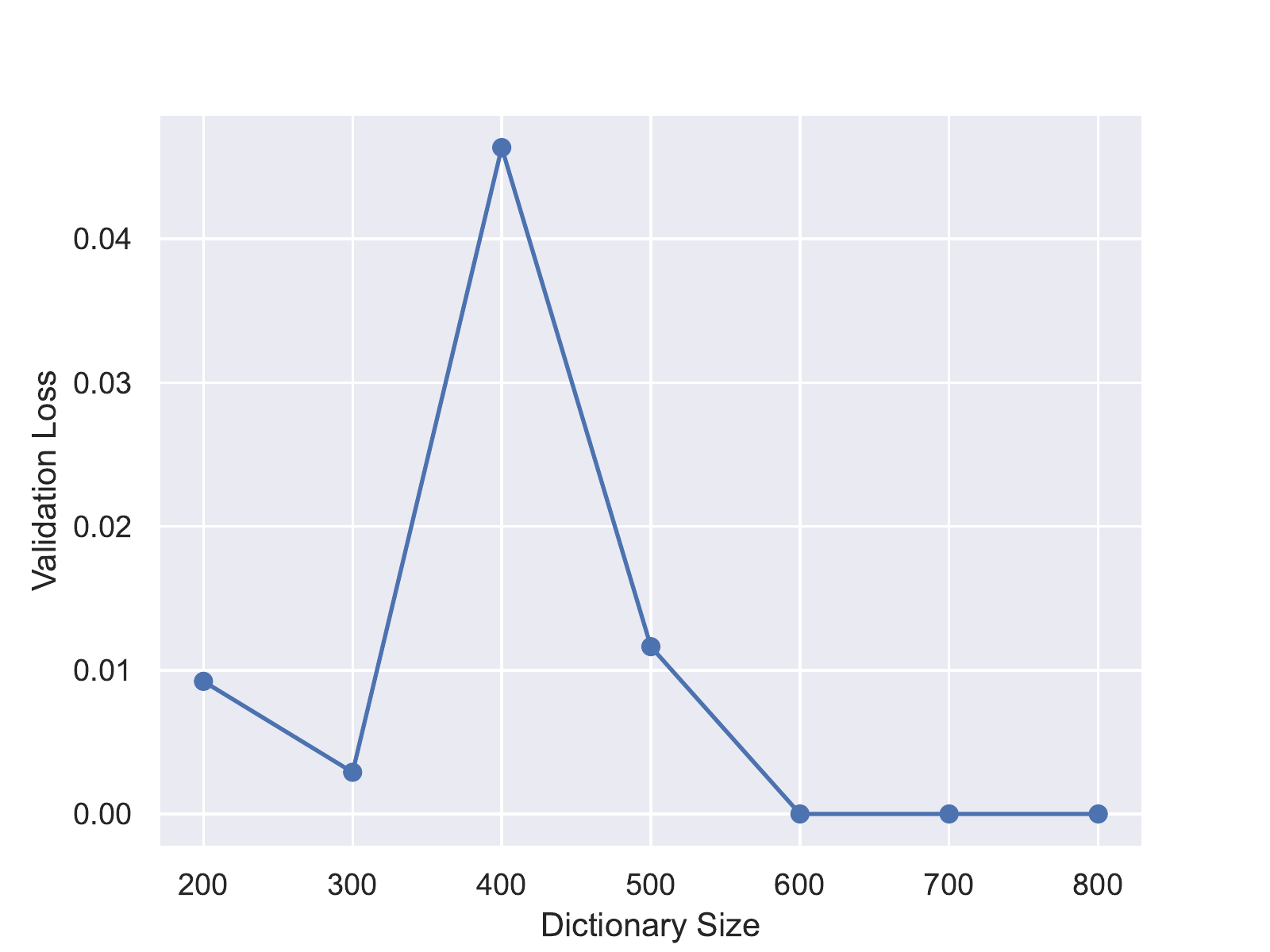} 

\caption{The results of overall MSE ($f + D + D^{*}$) on the validation set with different dictionary size. When the dictionary size is larger than 600, the loss on the validation set tends to be stable.} %

\smallskip

\label{Fig6} 
\end{figure}

(7)\textbf{ Effect of the dictionary size} was investigated based on the validation loss. The validation loss here was defined as the sum of loss of parameters $f$, $D$, and $D^{*}$ on the validation set. Results in Fig. \ref{Fig6} indicated that the validation loss decreased as the dictionary size changed from 200 to 300, but increased as dictionary size changed from 300 to 400. We went one step further to test dictionary sizes greater than 400 and found that validation loss further decreased and stayed nearly stable after 600. Because the training time increased dramatically as the dictionary size exceeds 600, e.g., the training time for an 800 dictionary was about twice that of a 600 dictionary, we deteremined the optimal dictionary size to be 600.

\begin{figure}[htbp] 
\centering 

\includegraphics[width=0.45\textwidth]{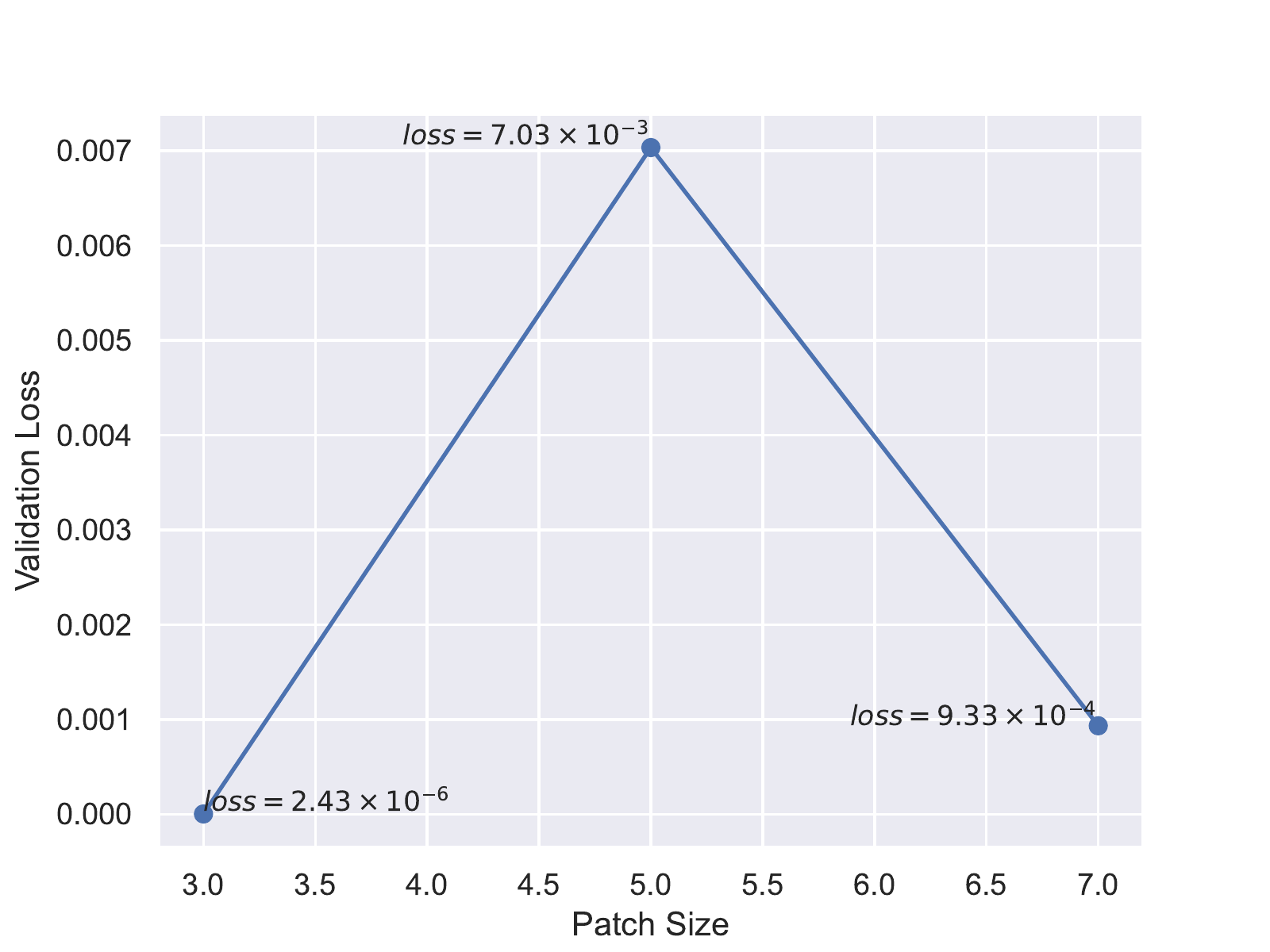} 

\caption{The impact of different patch sizes on the validation loss. The loss increases first and then decreases with the increase of patch size, and the minimal loss is achieved achieved at patch = 3.} %
\smallskip

\label{Fig7} 
\end{figure}

(8)\textbf{ Effect of the patch size} was also tested according to the validation loss. Fig. \ref{Fig7} showed a minimum loss at patch size of 3, compared to patch sizes of 5 and 7. Moreover, with the increase in patch size, training time increased sharply from 11h to 61h. 

From ablation experiments 1-8, we can conclude the optimal METSC structure consists of the Transformer encoder and the sparsity decoder with patched inputs at patch size of 3, using 5 b-values of  $20, 50, 150, 300, 500 s/mm^2$.

\subsection{Performance Evaluation}
We evaluated the performance of the optimized METSC network in terms of its robustness against noise (SNR), estimation accuracy compared to the other state-of-the-art algorithms, and its generalizability on multicenter data.

\begin{figure}[htbp] 
\centering 

\includegraphics[width=0.45\textwidth]{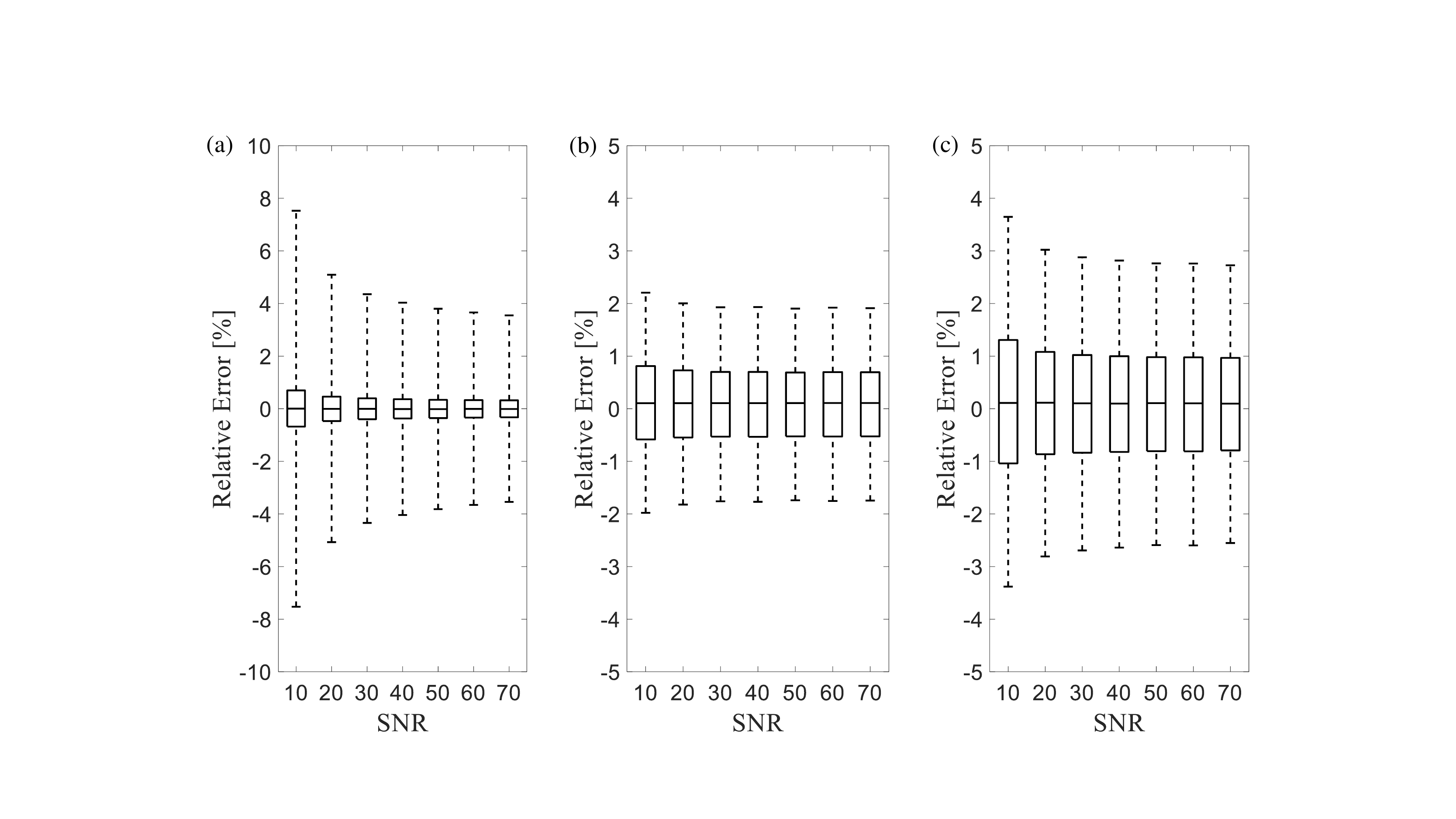} 

\caption{Relative errors (percentage of the gold standard at the different SNR levels) of the estimated IVIM parameters at SNR levels from 10 to 70.} %

\smallskip

\label{Fig8} 
\end{figure}

(1)\textbf{ Effects of SNR.} Different levels of noise were added into the signal according to Eq. [\ref{eq41}] resulting in SNR levels from 10 to 70, and we evaluated the relative error (percentage of the gold standard at the different SNR levels.) Fig. \ref{Fig8}(a) showed that the relative error of $f$ decreased gradually as SNR increased and stabilized at SNR above 40. In contrast, the estimation of $D$ was relatively insensitive to SNR (Fig. \ref{Fig8}(b)). The relative error of $D^{*}$ changed slightly with SNR and stabilized at SNR above 40. The results indicated the estimated parameters were relatively robust against noise and an SNR above 40 could ensure optimal accuracy.

(2)\textbf{ Comparison with other algorithms.} Four different algorithms were compared, including two optimization methods—the NLLS and Bayesian method ~\citep{gustafsson2018impact,jalnefjord2018comparison}, and three learning-based methods— q-DL ~\citep{golkov2016q}, IVIM-NET ~\citep{barbieri2020deep} and SCDNN ~\citep{TianshuZheng2021}. Compared with NLLS and Bayesian methods, the learning-based methods provided significantly higher estimation accuracy. Among the learning-based methods, the model-driven methods (SCDNN and METSC) outperformed the prior-information free q-DL and IVIM-Net, and the METSC demonstrated the best performance among the six methods (Table \ref{Tab 4}).
\begin{table}[!htb]
\centering
\caption{Comparison of six methods in estimating IVIM model parameters using a reduced number of b-values (5 b-values at $20, 50, 150, 300, 500$ $s/{mm}^{2}$). *p\textless 0.05, **p\textless 0.01,***p\textless 0.001 by paired t-test between each of the algorithms with METSC.}
{
\scalebox{0.6}{
\begin{tabular}{@{}ccccccc@{}}
\hline
                                         & \multicolumn{1}{c}{NLLS}           & \multicolumn{1}{c}{Bayesian}      & \multicolumn{1}{c}{{q}-DL}      & \multicolumn{1}{c}{IVIM-NET} & \multicolumn{1}{c}{SCDNN}        & \multicolumn{1}{c}{METSC}    \\ \hline
\multirow{1}{*}{{$f$}}  & \multirow{2}{*}{$0.54(***)$} & \multirow{2}{*}{$12(***)$} & \multirow{2}{*}{$0.7(***)$} & \multirow{2}{*}{$25(***)$} & \multirow{2}{*}{$0.072(*)$} & \multirow{2}{*}{$0.063$} \\
\multirow{1}{*}{$(\times 10^{-4})$}   & & & & & \\
\multirow{1}{*}{{$D $}}  & \multirow{2}{*}{$2.3\times{10}^{3}(***)$} & \multirow{2}{*}{$22(**)$}  & \multirow{2}{*}{$5.1(***)$} & \multirow{2}{*}{$2.8 \times 10^{3} (***)$} &\multirow{2}{*}{$3.1$}     & \multirow{2}{*}{$2.2$} \\
\multirow{1}{*}{$(\times 10^{-4} {\mu m}^{2}/ms)$} & & & & & \\
\multirow{1}{*}{{${D}^{*}$}} & \multirow{2}{*}{$19.8 \times{10}^{3} (***)$}    & \multirow{2}{*}{$42(***)$} & \multirow{2}{*}{$38(***)$} & \multirow{2}{*}{$1.3 \times 10^{3} (***)$} & \multirow{2}{*}{$1.8(*)$}  & \multirow{2}{*}{$1.4$} \\ 
\multirow{1}{*}{$(\times 10^{-2} {\mu m}^{2}/ms)$} & & & & & \\                  \hline

\end{tabular}}} \label{Tab 4}
\end{table}

(3) \textbf{Comparison with other algorithms in b-value choices.} To decouple the effects of the q-space sampling scheme and network performance, the three learning-based methods (q-DL, SCDNN, and METSC) were compared against the different b-value setups in the ablation experiment 6-7 in Section \ref{ablation experiments}. The results in Table \ref{Tab5} demonstrated that METSC achieved the best performance compared to other methods for all b-value combinations, and the optimal b-value choice was consistent with the ablation experiments (Table \ref{Tab3}).

\begin{figure}[!htb] 
\centering 

\includegraphics[width=0.47\textwidth]{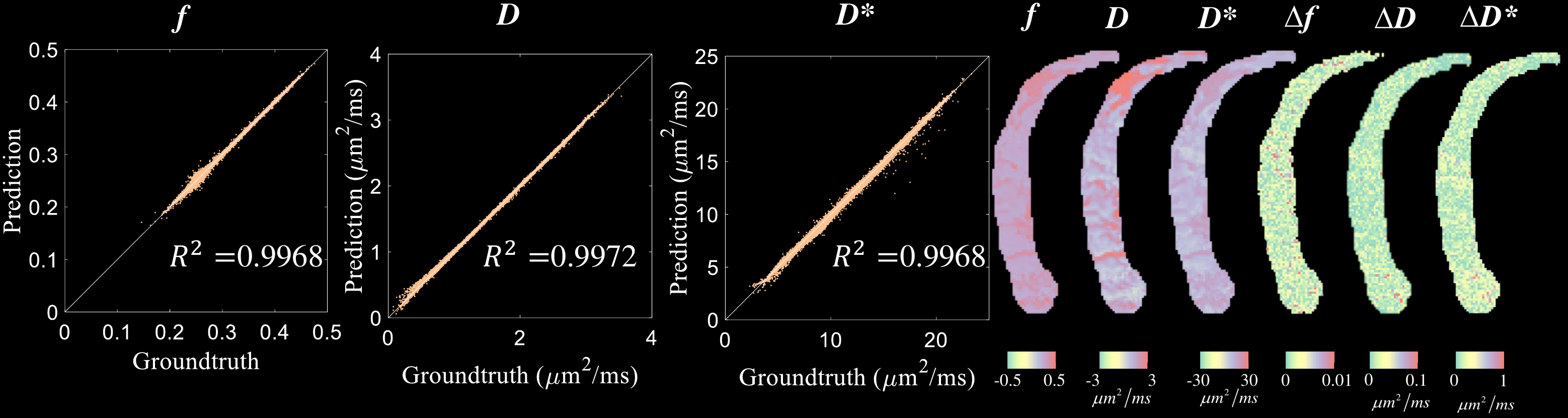} 

\caption{Estimated IVIM parameters in comparison with ground truth on the multicenter test data by METSC.} %

\smallskip

\label{Fig9}
\end{figure}

(4)\textbf{ Multicenter validation.} The previous tests were performed using training, validation, and testing data acquired on a 1.5T GE SIGNA HDXT scanner, and here we tested the network on data acquired on a 3.0T GE 750W scanner at another hospital with the same acquisition protocol. The new testing data included 2 patients (37194 voxels). The results in Table \ref{Tab6} showed that the METSC achieved the least estimation error using the reduced number of b-values (5 b-values at $20, 50, 150, 300, 500 s/mm^{2}$) compared to the other algorithms. The estimation accuracy was slightly reduced on the multicenter data compared to that on the single center, but still sufficient for parameter estimation, with $R^2$ between predicted values and ground truth over 0.996 (Fig. \ref{Fig9}). 

\begin{table}[!htb]
\centering
\caption{Evaluation of estimation errors on IVIM data acquired from another center with different scanner and field strength, using the five algorithms with five b-values.}
{
\scalebox{0.75}{
\begin{tabular}{@{}cccccc@{}}
\hline
                                         & \multicolumn{1}{c}{NLLS}           & \multicolumn{1}{c}{Bayesian}      & \multicolumn{1}{c}{{q}-DL}      & \multicolumn{1}{c}{SCDNN}        & \multicolumn{1}{c}{METSC}    \\ \hline
\multirow{1}{*}{{$f$}}  & \multirow{2}{*}{$320$} & \multirow{2}{*}{$30$} & \multirow{2}{*}{$29$} & \multirow{2}{*}{$4.4$} & \multirow{2}{*}{$0
076$}  \\
\multirow{1}{*}{$(\times 10^{-4})$}   & & & &  \\
\multirow{1}{*}{{$D $}}  & \multirow{2}{*}{$97$} & \multirow{2}{*}{$21$}  & \multirow{2}{*}{$12$} & \multirow{2}{*}{$0.72$} & \multirow{2}{*}{$0.027$} \\
\multirow{1}{*}{$(\times 10^{-4} {\mu m}^{2}/ms)$} & & & & & \\
\multirow{1}{*}{{${D}^{*}$}} & \multirow{2}{*}{$973$}    & \multirow{2}{*}{$45$} & \multirow{2}{*}{$34$} & \multirow{2}{*}{$8.4$}   & \multirow{2}{*}{$0.48$} \\ 
\multirow{1}{*}{$(\times 10^{-1} {\mu m}^{2}/ms)$} & & & & & \\                  \hline

\end{tabular}}} \label{Tab6}
\end{table}
\subsection{NODDI Model}
In this part, we first describe how we selected the public dataset and we compared our proposed framework with all other published methods for NODDI parameter estimation in terms of accuracy and precision. Also, we tested our framework with different q-space downsample schemes. 
\subsection{Dataset and Training}
The multi-shell dMRI from HCP data ~\citep{van2013wu} were acquired on a 3T MR scanner with 3 b-values ($b=1000, 2000, 3000 s/mm^{2}$) and 90 diffusion directions per b-value. We randomly selected 26 subjects and used 5 of them for training (with 10\% of the training samples as validation ~\citep{golkov2016q}), and 21 subjects for testing. To evaluate the proposed METSC, the dataset was downsampled to 30 gradient directions per b-shell at $b=1000$ and $b=2000 s/mm^{2}$ for comparison with other studies ~\citep{ye2017tissue}. The gold standard microstructural parameters were computed by the NODDI Matlab Tool Box ~\citep{zhang2012noddi} using all 270 q-space data. Similar to section 3.1, all datasets were split into overlapping patches with a step size of 1 in zero-padded images. 
\subsection{Performance Evaluation} 
The performance of the proposed network on NODDI was compared with six algorithms including the conventional dictionary learning based method AMICO ~\citep{daducci2015accelerated}, a a traditional q-space learning method q-DL ~\citep{golkov2016q}, two model-driven learning-based methods ~\citep{ye2017tissue,ye2020improved}. As MEDN ~\citep{ye2017estimation}, MEDN+ ~\citep{ye2017tissue}, MESC ~\citep{ye2019deep} and MESC-Net Sep\_Dict ~\citep{ye2020improved} (abbreviated as MESC\_Sep here) are variations of the same class of algorithm, and MEDN+ and MESC\_Sep have superior performance than MEDN and MESC, thus we only showed the results of MEDN+ and MESC\_Sep. We further tested the networks with number of diffusion directions. Robustness test was also carried out by smearing the signal and adding heavy noise.
\subsubsection{Comparison with other algorithms}
\label{noddi performance}
\begin{figure*}[!htb] 
\centering 

\includegraphics[width=0.8\textwidth]{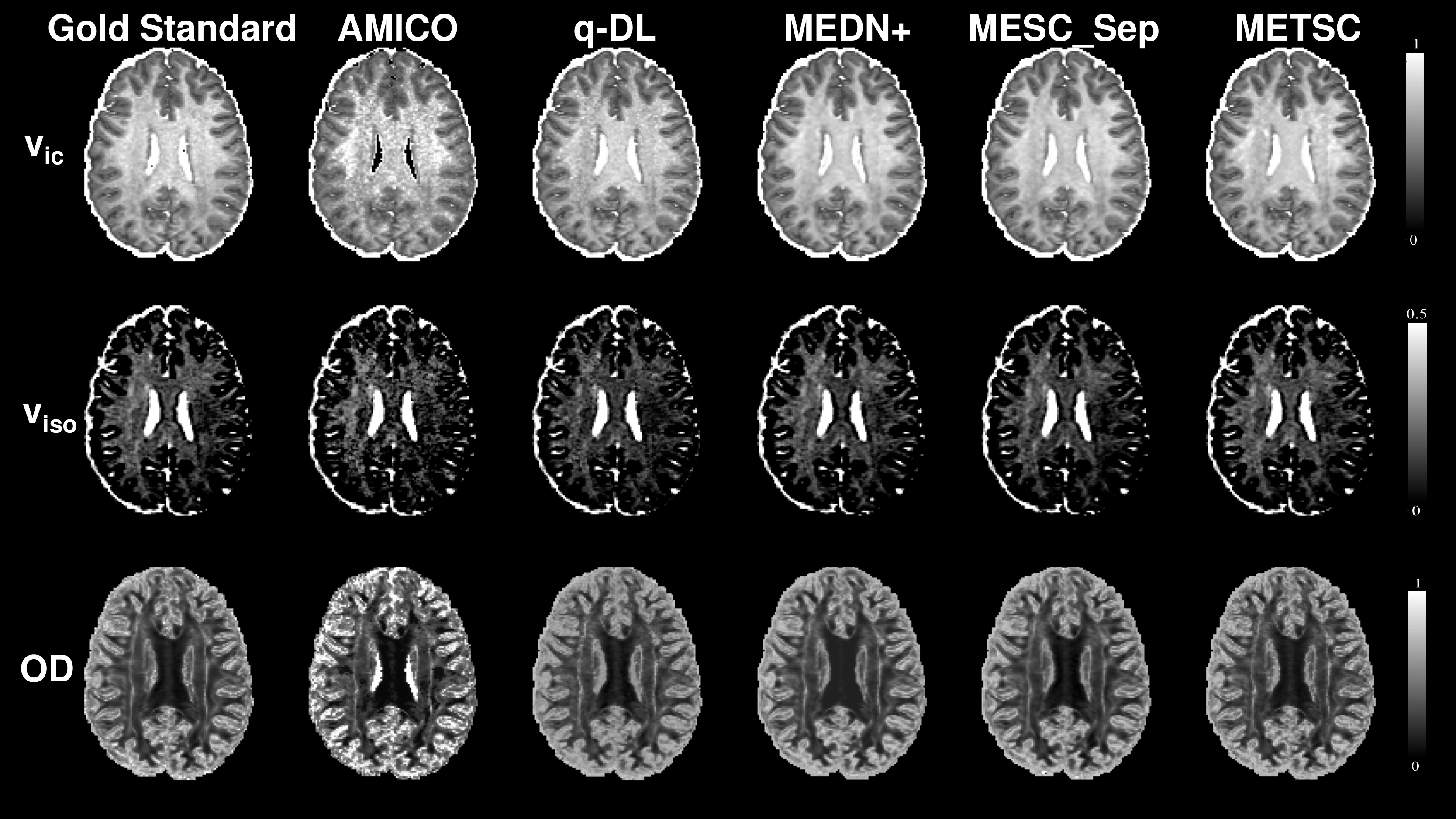} 

\caption{The gold standard and estimated NODDI parameter $v_{ic}$, $v_{iso}$, and OD based on AMICO, q-DL, MEDN+, MESC\_Sep, and METSC (ours) in a test subject with 30 diffusion directions per shell at b-valuse of 1000 and 2000 $s/mm^{2}$). } %

\smallskip

\label{Fig10} 
\end{figure*}

\begin{figure}[!htb] 
\centering 

\includegraphics[width=0.45\textwidth]{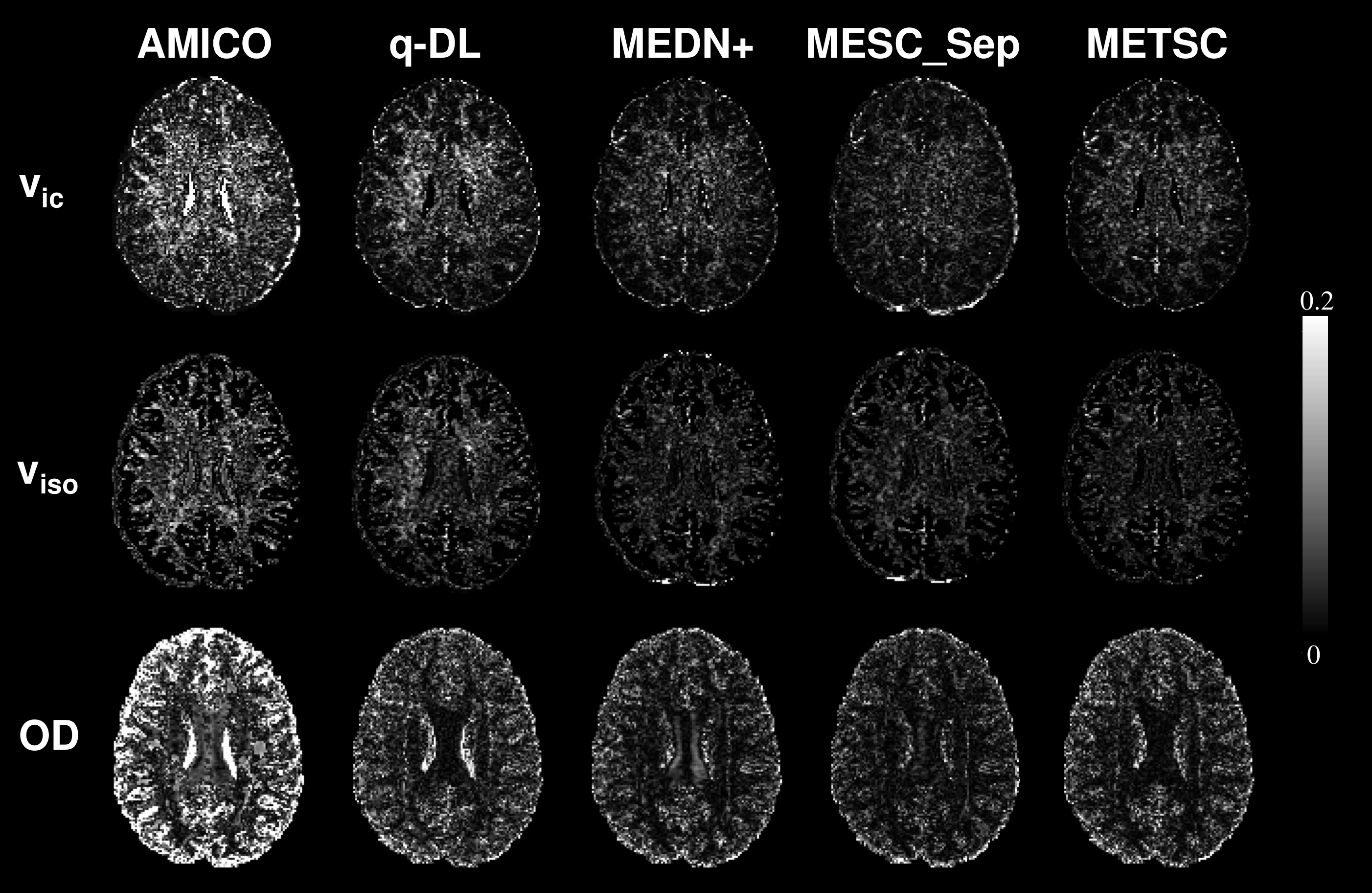} 

\caption{Estimation errors of $v_{ic}$, $v_{iso}$, and OD in a representative test subject using AMICO, q-DL, MEDN+, MESC\_Sep, and METSC (ours) in a test subject with 30 diffusion directions per shell at b-valuse of 1000 and 2000 $s/mm^{2}$). } %

\smallskip

\label{Fig11} 
\end{figure}

\begin{figure}[!htb] 
\centering 

\includegraphics[width=0.445\textwidth]{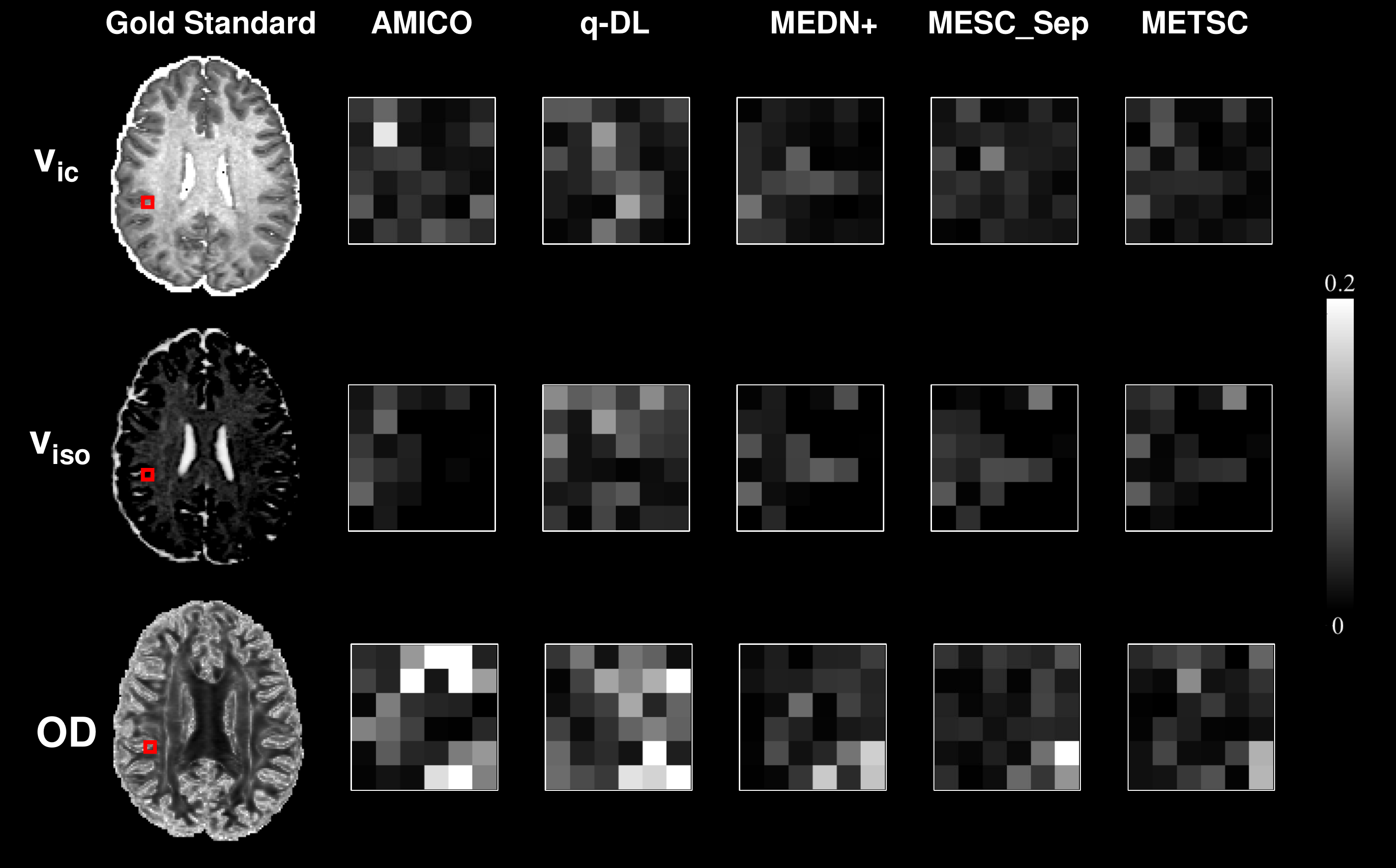} 

\caption{Zoom-in views of estimation errors of $v_{ic}$, $v_{iso}$, and OD in a representative test subject using AMICO, q-DL, MEDN+, MESC\_Sep, and METSC (ours) in a test subject with 30 diffusion directions per shell at b-valuse of 1000 and 2000 $s/mm^{2}$).} %

\smallskip

\label{Fig12} 
\end{figure}

In the estimation accuracy test, we used downsampled q-space with 30 diffusion directions per shell from 21 test subjects and compared the MSE of the different algorithms. Fig. \ref{Fig10} indicated that the $v_{ic}$, $v_{iso}$ and OD estimated from AMICO was overall worse than other learning-based methods. The proposed METSC provided the best results compared to other learning algorithms (Table \ref{Tab7}). The error maps between the were shown in Fig. \ref{Fig11}, and a zoomed view of error maps were illustrated in Fig. \ref{Fig12}. All results pointed to that the AMICO results were the worst with reduced q-space data, and our proposed METSC outperformed others. The mean and standard deviations of the average estimation errors across 21 test subjects were shown in Table \ref{Tab7} for all algorithms, which showed that the statistically reduced estimated errors by METSC lower than all other methods via paired Student’s t-test. 

\begin{table}[]
\caption{Evaluation of estimation errors on NODDI parameters using different methods on different number of diffusion directions. **p\textless0.01, ***p\textless0.001 by paired t-test between different methods with respect to the METSC is marked.} 
\centering
\scalebox{0.58}{%
\begin{tabular}{ccccccc}
\hline
                                                                                            &    & AMICO & q-DL & MEDN+ & MESC\_Sep & METSC \\ \hline
\multirow{6}{*}{\begin{tabular}[c]{@{}c@{}}30 diffusion gradients\\ per shell\end{tabular}} & \multirow{1}{*}{$v_{ic}$}  & \multirow{1}{*}{$4.9 \pm 1.8$}     & \multirow{1}{*}{$0.4 \pm 0.09$}   & \multirow{1}{*}{$0.3 \pm 0.04$}     & \multirow{1}{*}{$7.0 \pm 2.8$}         & \multirow{2}{*}{$0.08 \pm 0.01$}     \\
                                                                                            & $\times 10^{-2}$  & \multirow{1}{*}{(***)}     & \multirow{1}{*}{(***)}    &   \multirow{1}{*}{(**)}   & \multirow{1}{*}{(***)}        &     \\
                                                                                            & $v_{iso}$  & \multirow{1}{*}{$3.0 \pm 1.3$}     & \multirow{1}{*}{$1.7 \pm 0.2$}    & \multirow{1}{*}{$1.5 \pm 0.3$}     & \multirow{1}{*}{$1.8 \pm 0.4$}         & \multirow{2}{*}{$0.4 \pm 0.02$}     \\
                                                                                            & $\times 10^{-3}$  & \multirow{1}{*}{(***)}     & \multirow{1}{*}{(***)}    & \multirow{1}{*}{(***)}     & \multirow{1}{*}{(***)}         &      \\
                                                                                            & $OD$  &  \multirow{1}{*}{$31 \pm 29$}     &  \multirow{1}{*}{$4.2 \pm 0.3$}    &  \multirow{1}{*}{$3.9 \pm 0.2$}     &  \multirow{1}{*}{$2.3 \pm 2.3$}         &  \multirow{2}{*}{$0.9 \pm 0.01$}     \\
                                                                                            & $\times 10^{-3}$  & \multirow{1}{*}{(***)}     & \multirow{1}{*}{(***)}   & \multirow{1}{*}{(***)}     & \multirow{1}{*}{(***)}         &      \\ \hline
\multirow{6}{*}{\begin{tabular}[c]{@{}c@{}}18 diffusion gradients\\ per shell\end{tabular}} & $v_{ic}$  & \multirow{1}{*}{$6.7 \pm 2.0$}     & \multirow{1}{*}{$0.5 \pm 0.1$}    & \multirow{1}{*}{$0.4 \pm 0.03$}     & \multirow{1}{*}{$8.9 \pm 8.7$}         & \multirow{2}{*}{$0.09 \pm 0.01$}     \\
                                                                                            & $\times 10^{-2}$  & \multirow{1}{*}{(***)}     & \multirow{1}{*}{(***)}    & \multirow{1}{*}{(***)}     & \multirow{1}{*}{(***)}        &      \\
                                                                                            & $v_{iso}$  & \multirow{1}{*}{$5.7 \pm 1.4$}     & \multirow{1}{*}{$2.9 \pm 1.5$}    & \multirow{1}{*}{$2.0 \pm 0.8$}     & \multirow{1}{*}{$2.2 \pm 0.5$}         & \multirow{2}{*}{$0.5 \pm 0.03$}     \\
                                                                                            & $\times 10^{-3}$  & \multirow{1}{*}{(***)}     & \multirow{1}{*}{(***)}   & \multirow{1}{*}{(***)}     &       \multirow{1}{*}{(***)}   &      \\
                                                                                            & $OD$  & \multirow{1}{*}{$54 \pm 31$}     & \multirow{1}{*}{$7.8 \pm 1.3$}   & \multirow{1}{*}{$4.9 \pm 1.3$}     & \multirow{1}{*}{$11 \pm 11$}         & \multirow{2}{*}{$1.5 \pm 0.02$}     \\
                                                                                            & $\times 10^{-3}$  & \multirow{1}{*}{(***)}     & \multirow{1}{*}{(***)}    & \multirow{1}{*}{(***)}     & \multirow{1}{*}{(***)}         &      \\ \hline
\multirow{6}{*}{\begin{tabular}[c]{@{}c@{}}12 diffusion gradients \\ per shell\end{tabular}} & $v_{ic}$  & \multirow{1}{*}{$7.5 \pm 2.0$}     & \multirow{1}{*}{$1.0 \pm 0.1$}    & \multirow{1}{*}{$0.6 \pm 0.3$}     & \multirow{1}{*}{$51 \pm 9.2$}         & \multirow{1}{*}{$0.1 \pm 0.02$}     \\
                                                                                            & $\times 10^{-2}$  & \multirow{1}{*}{(***)}     & \multirow{1}{*}{(***)}    &    \multirow{1}{*}{(***)}    & \multirow{1}{*}{(***)}         &      \\
                                                                                            & $v_{iso}$  & \multirow{1}{*}{$7.4 \pm 2.4$}     & \multirow{1}{*}{$7.3 \pm 1.5$}    & \multirow{1}{*}{$3.5 \pm 0.8$}     & \multirow{1}{*}{$2.6 \pm 0.5$}         & \multirow{2}{*}{$0.6 \pm 0.03$}     \\
                                                                                            & $\times 10^{-3}$  & \multirow{1}{*}{(***)}     & \multirow{1}{*}{(***)}    & \multirow{1}{*}{(***)}     &      \multirow{1}{*}{(***)}      &      \\
                                                                                            & $OD$  & \multirow{1}{*}{$87 \pm 31$}     & \multirow{1}{*}{$8.9 \pm 1.3$}    & \multirow{1}{*}{$5.1 \pm 0.4$}     & \multirow{1}{*}{$60 \pm 7.0$}         & \multirow{2}{*}{$2.0 \pm 0.2$}     \\
                                                                                            & $\times 10^{-3}$  & \multirow{1}{*}{(***)}     & \multirow{1}{*}{(***)}    &    \multirow{1}{*}{(**)}    & \multirow{1}{*}{(***)}         &      \\ \hline
\end{tabular}%
\label{Tab7}

}
\end{table}
\subsubsection{Effect of the different number of diffusion directions}
In the previous experiments, the number of diffusion directions was set to be 30 for each shell ($b=1000, 2000 s/mm^{2}$). In this part, we further reduced the number of directions to 18 and 12 for each shell. The results in Table \ref{Tab7} demonstrated METSC achieved minimal estimation errors compared to other algorithms for all choices of gradient numbers.
\subsubsection{Robustness test}
Beyond evaluating the estimation accuracy as done in previous studies ~\citep{golkov2016q,ye2017estimation,ye2017tissue,ye2019deep,ye2020improved}, this study also investigated the robustness of the network. The robustness test was divided into two parts, the first part tested how the choice of diffusion directions affected the results ~\citep{karimi2021deep}, and the second part tested the robustness of the network in response to the abnormal input signal. 

\begin{figure}[!htb] 
\centering 

\includegraphics[width=0.445\textwidth]{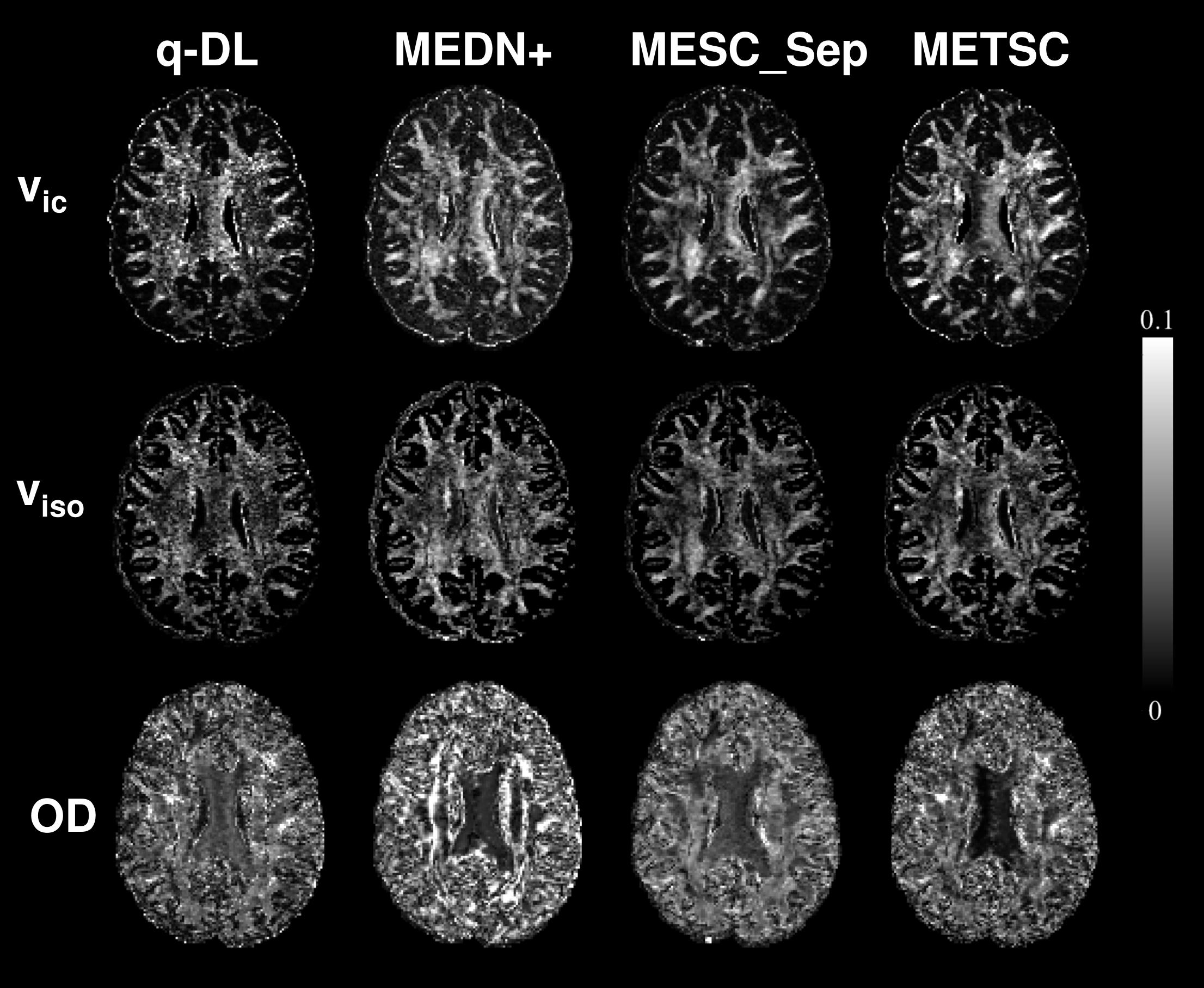} 

\caption{Evaluation of estimation precision on four learning-based algorithms. The maps showed the standard deviation of estimated parameters between bootstrapped results with shuffled diffusion direcitons. The METSC (ours) method showed higher precision than the other model-based methods MEDN+, and MESC\_Sep, indicated by the lower standard deviation from bootstrap. } %
 
\smallskip

\label{Fig13} 
\end{figure}

\begin{figure}[!htb] 
\centering 

\includegraphics[width=0.45\textwidth]{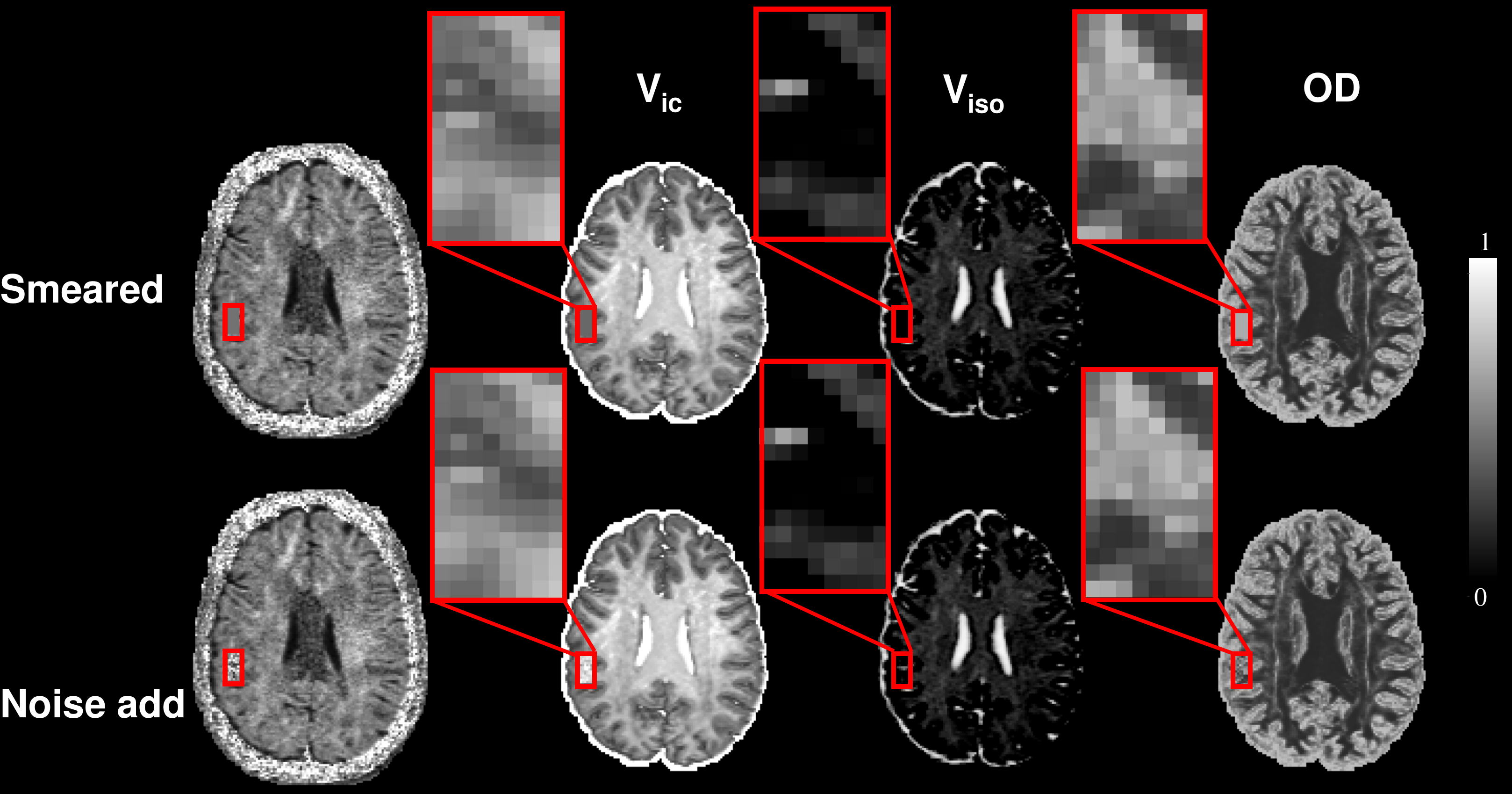} 

\caption{Evaluation of robustness by testing whether the METSC (ours) method will generate fake output that is not supported by the data. The signal patch in the red rectangle is smeared (a) or noised corrupted (b). Both results demonstrated the network will not produce the unsupported estimation. } %

\smallskip

\label{Fig14} 
\end{figure}

In the first test, we used three different combinations of diffusion directions (n=30) by bootstrap, and the standard deviation of estimated parameters from the three datasets was used to evaluate the robustness. Fig. \ref{Fig13} indicated that METSC resulted the least variation among the bootstraps and thus the highest robustness. When the network took the abnormal inputs, such as the smeared input and the noise added input (Fig. \ref{Fig14}), it did not generate unexpected / forged outputs.

\section{Discussion}
In this study, a model-bias was introduced to facilitate the training of the transformer structure. And in this part, the discussion will be talked about our motivation, how the model-bias work, how we setup our network, and our future work.
\subsection{Motivation}
Motivated by the feature extraction capacity of the Transformer, in this work, we proposed a \emph{Microstructure Estimation Transformer with Sparse Coding} (METSC) that integrates the Transformer encoder with a model-based sparsity decoder to enhance the model estimation and enabled the Transformer to be efficiently trained with limited data. To our best knowledge, this is the first time the Transformer structure is applied to a regression task in the medical imaging area, especially for dMRI-based microstructural parameter estimation. Meanwhile, it is also the first time a physiological model bias is introduced into the Transformer structure via an iterative optimization technique, which not only improves the model interpretability but also the training efficiency of the Transformer. 

To demonstrated the generalizability of METSC, we chose the IVIM and NODDI models to testing the network performance as they are representative of various types of dMRI models, e.g., IVIM model is a clinically useful model that estimates microcapillary flow from multiple b-values and NODDI emphasizes high angular resolution for resolving neurite orientation and is heavily studied with deep learning techniques. By modifying stage 3 in Fig. \ref{Figure1} to a specific model configuration, the suggested METSC framework can be adapted to a variety of signal models beyond dMRI models, such as T1 and T2 mapping.
\subsection{Model-bias}
A key contribution of this work is that we brought up and validated the hypothesis that the model-bias can partially solve the data-hungry of the ViT. Compared with the ViT structure, incorporating the model-bias into METSC significantly improved its performance with the small amount of training samples. As demonstrated in the experiments (Section \ref{ablation experiments} and Section \ref{noddi performance}), the proposed METSC framework was compared with the other networks, with no more than 0.3M training samples in the IVIM model and 1.5M in the NODDI model. Because the nonlinearity of NODDI is much higher than the IVIM model, it is expected that NODDI needs more data to learn the dictionary. 

\begin{figure}[htbp] 
\centering

\includegraphics[width=0.45\textwidth]{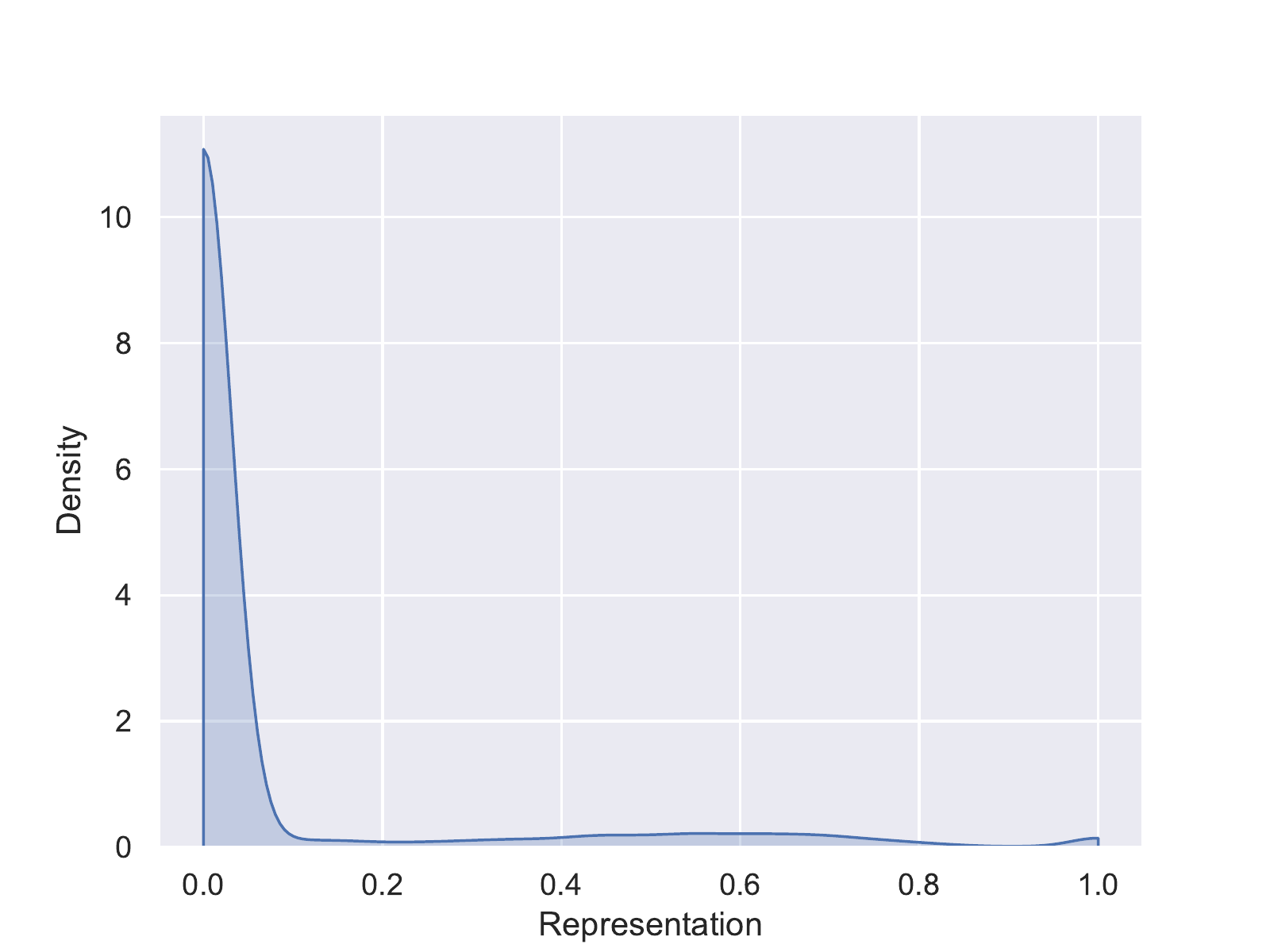} 

\caption{The distribution of dictionary coefficients in the sparse representation $\boldsymbol{x}$ given by METSC for a representative test subject, indicaing high sparsity in the dictionary coefficients.} %

\smallskip

\label{Fig15} 
\end{figure}

Meanwhile, since the model-bias is a kind of sparse representation, the network could be examined by the sparsity of the representation signals. For instance, using the NODDI model, Fig. \ref{Fig15} showed the distribution of nonzero entries in the dictionary coefficients for a test subject and the sparsity was over 84\%, which supported the validity of our sparsity hypothesis.
\subsection{Network setup}
To access the architecture of the METSC framework, various ablation experiments on hyper-parameters and the structure of METSC were investigated. The METSC framework contains a large number of hyper-parameters. Therefore, we only focused on the major hyper-parameters in the iterative decoder phase, while the hyper-parameters in the Transformer-based encoder were fixed as~\citep{dosovitskiy2020image}. We further investigated the encoder and decoder performance with respect to input patch size and dictionary size, which were not fully addressed in previous studies. In terms of data input, we explored the effect of patch size, that has not been investigated before and determined an optimal patch size of 3 could gave the lowest loss and best computational efficiency. We also tested the dictionary size N of the decoder on the IVIM model and found that the results for N $ \textless$ 400 were the same as ~\citep{ye2020improved}, which however, did not explore N $\textgreater$ 400. In this work, we tested the whole spectrum of dictionary size from 200-800 and determined dictionary size of 600 was optimal for both accuracy and efficiency. 
\begin{figure*}[!htb] 
\centering 

\includegraphics[width=0.72\textwidth]{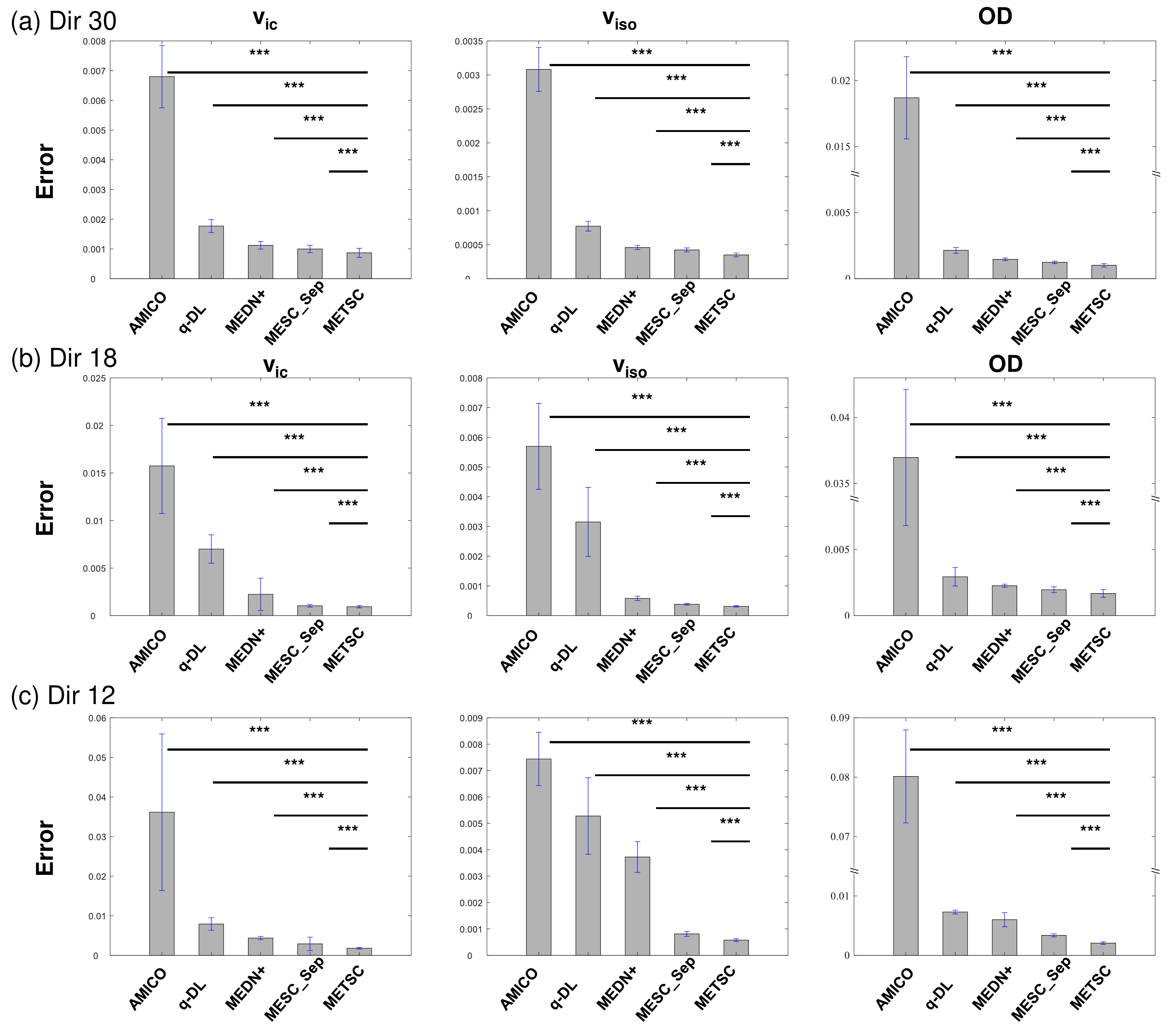} 

\caption{The means and standard deviations of the whole-brain average estimation errors of NODDI parameters in test subjects (n=21) using different estimation algorithms at downsampled diffusion directions of 30, 18, and 12 per shell at b-values of 1000 and 2000 $s/mm^2$. *p\textless0.05, **p\textless0.01, ***p\textless0.001 by paired t-test.} %

\smallskip

\label{Fig16} 
\end{figure*}

Note that in the NODDI experiments, compared with the previous model-based network MESC\_Sep ~\citep{ye2020improved}, METSC did not use LSTM to incorporate the historical information. Although incorporating historical information can improve the accuracy of estimation, we found it made the network unstable e.g., a slight change in the diffusion direction may reduce the fitting accuracy and it is known the preprocessing steps of registration and motion correction could easily change the actual gradients. This can be partially compensated by interpolating the diffusion directions to match the target dataset ~\citep{qin2020knowledge}. The results with the interpolated diffusion directions were shown in Fig \ref{Fig16}(a~c), which showed the performance of MESC\_Sep after interpolation on shore basis ~\citep{merlet2013continuous} were improved with this strategy but still cannot beat METSC. Combining the IVIM and NODDI results in Fig. \ref{Fig13}, Fig. \ref{Fig16}, Table \ref{Tab5}, and Table \ref{Tab7}, it can be concluded that our proposed method has the best robustness against both the directions and magnitudes (b-value) of diffusion gradients.

\subsection{Future work}

Training of the learning-based network requires densely sampled diffusion signals with high quality ~\citep{golkov2016q,ye2017estimation,ye2017tissue,ye2019deep,ye2020improved,chen2020estimating}, which are often difficult to obtain in practice. Recent studies have demonstrated the pre-trained model can help the microstructural estimation using an auxiliary dataset ~\citep{li2021pretraining}. Similar the great success of the pre-trained Transformer model in NLP ~\citep{brown2020language,devlin2018bert,radford2018improving}, our pre-trained METSC can probably transfer the knowledge to other domains that are not limited to the dMRI models but also T1 mapping, T2 mapping, other multi-pool models, which will be investigated in future work. 
\section{Summary and Conclusion }
In this work, we proposed a novel model-driven Transformer with sparse coding to estimate microstructural parameters in dMRI models with reduced q-space data. The proposed METSC framework integrated the strength of the Transformer and also address the large training data requirement of ViT by introducing model-based inductive bias. Compared with the conventional optimization methods and the state-of-the-art learning-based methods, METSC achieved the highest accuracy in estimating model parameters for both IVIM and NODDI models. The network also showed good interpretability, generalizability and robustness, and thus, is potentially useful for fast dMRI acquisition with undersampled q-space, which may be particularly important for dMRI of the moving subjects.

\section*{Acknowledgments}
This work is supported by Ministry of Science and Technology of the People’s Republic of China (2018YFE0114600), National Natural Science Foundation of China (61801424, 81971606, 82122032), and Science and Technology Department of Zhejiang Province (202006140, 2022C03057).

\section*{Appendix}

\renewcommand{\thefigure}{A\arabic{figure}}
\renewcommand{\thetable}{A\arabic{table}}
\setcounter{table}{0}
The following Appendix describes the comparison of three supervised learning based methods in estimating IVIM model parameters using different combinations of b-values, in terms of MSE.
\begin{table}[!htb]
\centering
\caption{Comparison of three supervised learning based methods in estimating IVIM model parameters using different combinations of b-values, in terms of MSE. The combinations with top performance were highlighted in bold.}
\resizebox{3.45in}{!}{%
\begin{tabular}{ccccc}
\hline
                                  &       & $f  $     & $D  $    & $D^{*} $   \\ 
\multicolumn{1}{l}{} & \multicolumn{1}{l}{} & \multicolumn{1}{l}{($\times 10^{-4})$} & \multicolumn{1}{l}{$(\times 10^{-4} {\mu m}^{2}/ms)$} & \multicolumn{1}{l}{$(\times 10^{-2} {\mu m}^{2}/ms)$} \\ \hline

3 b-values        & q-DL  & 2.3   & 12  & 27  \\
(20, 150, 500) $s/mm^2$                                  & SCDNN & 2     & 37  & 14  \\
                                  & METSC & \textbf{0.41}  & \textbf{2.8} & \textbf{8.4} \\ \hline
5 b-values Comb1 & q-DL  & 0.68  & 5.1 & 38  \\
(20, 50, 150, 300, 500) $s/mm^2$                                   & SCDNN & 0.072 & 3.1 & 1.8 \\
                                  & METSC & \textbf{0.063} & \textbf{2.2} & \textbf{1.4} \\
5 b-values Comb2 & q-DL  & 4     & 4.6 & 22  \\
(20, 50, 150, 200, 500) $s/mm^2$                                  & SCDNN & 7.1   & 2.1 & 6.2 \\
                                  & METSC & \textbf{0.14}  & \textbf{2.1} & \textbf{1.8} \\
5 b-values Comb3 & q-DL  & 0.39  & 8.8 & 45  \\
(20, 50, 200, 300, 500)  $s/mm^2$                                 & SCDNN & 0.4   & 2.1 & 9   \\
                                  & METSC & \textbf{0.063} & \textbf{5.6} & \textbf{1.7} \\
5 b-values Comb4 & q-DL  & 0.43  & 8.8 & 46  \\
(20, 100, 150, 300, 500)  $s/mm^2$                                  & SCDNN & 0.27  & 9   & 2.2 \\
                                  & METSC & \textbf{0.097} & \textbf{1.7} & \textbf{1.7} \\
5 b-values Comb2 & q-DL  & 4.1   & 5   & 29  \\
(20, 50, 150, 200, 500)  $s/mm^2$                                & SCDNN & 0.19  & 3.2 & 2.7 \\
                                  & METSC & \textbf{0.094} & \textbf{1.7} & \textbf{2}   \\ \hline
7 b-values       & q-DL  & 0.37  & 9.7 & 2.8 \\
(20, 50, 100, 150, 200,                                & SCDNN & 0.1   & 3.2 & 2.8 \\
   300, 500)   $s/mm^2$                               & METSC & \textbf{0.086} & \textbf{1.4} & \textbf{2.6} \\ \hline
\end{tabular}%
} \label{Tab5}
\end{table}

\bibliographystyle{model2-names.bst}\biboptions{authoryear}
\bibliography{ref}

\end{document}